\documentclass[twocolumn,groupedaddress,aps,pra,
amsmath,amssymb,showpacs]{revtex4-1}
\usepackage[colorlinks,linkcolor=blue,anchorcolor=blue,
            citecolor=blue,bookmarksnumbered]{hyperref}
\usepackage{graphicx,epsfig,float,epstopdf}
\usepackage{txfonts}
\usepackage{amssymb,amsmath,subeqnarray,mathrsfs,
           amsfonts,makecell,cases,color,extpfeil,lipsum}
\usepackage{comment,bm}
\allowdisplaybreaks

\begin{document}
\title{Simultaneous Quantum Squeezing of Light Polarizations and Atomic Spins in a Cold Atomic Gas}
\author{Jinzhong Zhu$^{1}$}
\author{Yue Mu$^{1}$}
\author{Guoxiang Huang$^{1,2,3,}$}\thanks{gxhuang@phy.ecnu.edu.cn}
\affiliation{$^1$State Key Laboratory of Precision Spectroscopy,
             East China Normal University, Shanghai 200241, China\\
             $^2$NYU-ECNU Institute of Physics, New York University at Shanghai, Shanghai 200062, China\\
             $^3$Collaborative Innovation Center of Extreme Optics, Shanxi University, Taiyuan, Shanxi 030006, China
             }
\begin{abstract}
We present a scheme to realize simultaneous quantum squeezing of light polarizations and atomic spins via a perturbed double electromagnetically induced transparency (DEIT) in a cold four-level atomic ensemble coupled with a probe laser pulse of two polarization components.
We derive two coupled quantum nonlinear Schr\"odinger equations from Maxwell-Heisenberg-Langevin equations describing the quantum dynamics of the atoms and the probe pulse, and develop a quantum theory of vector optical soliton (VOS), which have ultraslow propagation velocity and extremely low generation power. We solve the non-Hermitian eigenvalue problem describing the quantum fluctuations on the background of the VOS, and rigorously prove that all fluctuation eigenmodes (including continuous modes and four zero modes) obtained constitute a bi-orthonormal and complete set. We find that, due to the giant self- and cross-Kerr nonlinearities contributed by the perturbed DEIT, a large polarization squeezing of the probe pulse can be realized. We also find that, together with the polarization squeezing of the probe pulses, a significant squeezing of atomic spins also occurs simultaneously. The results of the simultaneous squeezing of light polarizations and atomic spins by using only a coherent probe pulse reported here opens a route for uncovering the unique property of the quantum interface between light and atomic ensembles, and also for applications in quantum information and precision measurement.
\end{abstract}

\maketitle

\section{Introduction}


Quantum squeezing, by which quantum fluctuations in one physical observable are reduced below standard quantum limit, at the expense of increased fluctuations in the corresponding conjugate observable, belongs to one of the most prominent nonclassical resources. It has compelling applications for quantum enhanced metrology (e.g., for the improvement of the sensitivity of gravitational-wave detectors) and is quite useful for the fundamental study on the physics of quantum entanglement. In recent years, the research for seeking efficient methods to create quantum squeezing has been at heart of the modern development of quantum optics~\cite{Drummond2004,Schnabel2017}.

Among many media for creating light squeezing (including parametric down-conversion crystals, optical fibers, and semiconductor lasers, etc.)~\cite{Andersen2016}, atomic gas is the one by which the squeezed light
was first realized experimentally, reported
by Slusher {\it et al.} in 1985~\cite{Slusher1985}. Later on, in order to overcome detrimental Raman scattering and fluorescence occurring with four-wave-mixing process, twin beam squeezing technique was used to improve light squeezing in double $\Lambda$-shaped atomic gases~\cite{Lett2007,Lett2008,Glorieux2011}. Polarization squeezing of light in an atomic vapor by using Faraday rotation was also considered~\cite{Ries2003,Barreiro2011}.

The study on the generation of light squeezed states stimulated many efforts to search for similar nonclassical states, i.e. spin squeezed states, in atomic ensembles~\cite{Kitagawa1993,Wineland1994}, which can be used to suppress uncertainties resulted from quantum fluctuations in spins associated with atomic states and has promising practical applications (e.g., improving long-term stability of  optical atomic clocks~\cite{Andre2004,Pedrozo2020,Schulte2020}).
Up to now,
spin squeezing has been realized by using direct interaction between atoms (such as Bose-Einstein condensates of atomic gases~\cite{Ma2011,Gross2012,Sinatra2022}), by mapping squeezed light onto atoms and employing quantum nondemolition measurements, etc.~\cite{Hald1999,Kuzmich2000,Julsgaard2001,Dantan2003,Akamatsu2004,
Dantan2004,Takeuchi2005,Dantan2006,Appel2008,Honda2008,Appel2009,Takano2009,
Schleier2010,Leroux2010,Hammerer2010,Bohnet2014,Polzik2016,Braverman2019,
XiaoYH2020}.

In recent studies~\cite{Zhu2021,Zhu2022}, it has been shown that light squeezing can be realized in a three-level atomic gas working on the condition of a perturbed electromagnetically induced transparency (EIT)~\cite{Fleischhauer2005}.
By introducing a non-zero but small two-photon detuning (which makes the system to deviate strict EIT condition slightly), the system supports giant optical Kerr nonlinearity and displays second-order dispersion effect, thereby
allows the formation of (single-component) ultraslow weak-light solitons with very low loss~\cite{Deng2004PRL,Huang2005PRE}. Due to the existence of the giant Kerr nonlinearity, the quantum squeezing of the slow-light solitons can be obtained~\cite{Zhu2021,Zhu2022}.


In this article, we present a scheme to realize simultaneous quantum squeezing of light polarizations and atomic spins via a perturbed double EIT (DEIT) in a cold four-level atomic ensemble coupled with a probe laser pulse with two polarization components. We derive two coupled (two-component) quantum nonlinear Schr\"odinger (NLS) equations from the Maxwell-Heisenberg-Langevin (MHL) equations controlling the quantum dynamics of the atoms and the probe pulse and develop a quantum perturbation theory of vector optical soliton (VOS)~\cite{Kivshar2003}. Such a VOS has ultraslow propagation velocity and extremely low generation power, contributed by the DEIT effect.

To investigate the quantum fluctuations on the background of the VOS,
we solve the non-Hermitian eigenvalue problem describing the quantum fluctuations
and present the solution for all eigenmodes, which include continuous modes and four zero modes. We prove {\it rigorously} that these eigenmodes constitute a bi-orthonormal and complete set and hence they provide an expansion basis for all possible quantum fluctuations.  Based on the giant self- and cross-Kerr nonlinearities resulted from the perturbed DEIT, we demonstrate that a large polarization squeezing of the probe field can be realized by inputting a coherent probe pulse. Furthermore, we also demonstrate that, together with the polarization squeezing of the probe pulse, a significant squeezing of atomic spins can occur {\it simultaneously}. We find that the zero modes of the quantum fluctuations play key roles for the quantum squeezing of the light polarization and the atomic spins in the system.

We stress that, although a multitude of researches on polarization squeezing of light and spin squeezing of atoms~\cite{Ma2011,Gross2012,Sinatra2022,Hald1999,Kuzmich2000,Julsgaard2001,Dantan2003,Akamatsu2004,
Dantan2004,Takeuchi2005,Dantan2006,Appel2008,Honda2008,Appel2009,Takano2009,
Schleier2010,Leroux2010,Hammerer2010,Bohnet2014,Polzik2016,Braverman2019,
XiaoYH2020,Chirkin1993,Korolkova2002,Bowen2003,Heersink2003,Josse2003,Heersink2005,
Luis2006,Sherson 2006,Wolfgramm2010,Wen2017}, as well as vector solitons~\cite{Chen1995,Lee2004,Rand2005-1,Rand2005-2,Corney2006,Tsang2006,Kravtsov2007,
Corney2008,Milanovic2010,Sorokin2021} in various systems have been studied in the past years, our work is different from them, including the physical model, squeezing mechanism, theoretical method, and research results. Since
the possibility to generate the simultaneous squeezing of light polarization and atomic spins by using only an input of coherent-state probe pulse has never been explored before, the research result reported here paves a way not only for revealing the unique property of the quantum interface between light and atomic ensembles, but also for promising applications in quantum information processing and precision measurement.

The remainder of the article is organized as follows.
In Sec.~\ref{Sec2}, we present the physical model under study and derive the coupled quantum NLS equations describing the nonlinear propagation of the two polarization components of the probe pulse.
In Sec.~\ref{Sec3}, we solve the non-Hermitian eigenvalue problem for the quantum fluctuations around the VOS, provide its solution for the all fluctuation eigenmodes, and  prove their bi-orthonormality and completeness rigorously.
In Sec.~\ref{Sec4}, we study the quantum dynamics of the VOS, the
polarization squeezing of the probe pulse, and the spin squeezing of the atoms. A summary of the main results obtained in this work is given in Sec.~\ref{Sec5}.  Some calculation details omitted in the main text are provided in four appendices.

\section{Model and coupled quantum NLS equations}\label{Sec2}

\subsection{Physical model}\label{Sec2A}

We consider a cold atomic gas with a tripod-shaped four-level configuration~\cite{Petrosyan2004}, interacting with a weak, pulsed probe laser field  $\hat{\bf E}_p$ (with central angular frequency $\omega_{p}$ and wavenumber $k_p=\omega_{p}/c$) and a strong, continuous-wave control laser field ${\bf E}_c$
(with central angular frequency $\omega_{c}$ and wavenumber $k_c=\omega_{c}/c$).
\begin{figure}
\centering
\includegraphics[width=1\columnwidth]{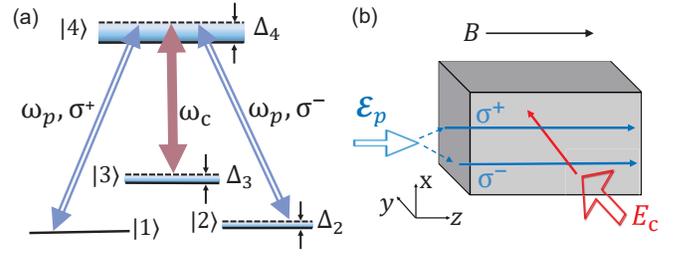}
\caption{
(a)~Energy-level diagram and DEIT  excitation scheme of the atomic gas with the tripod-type four-level configuration. $\omega_{p}$ is the central angular frequency of the quantized probe pulse, which can be taken as a linear superposition of the right-circularly ($\sigma^{+}$) and the left-circularly ($\sigma^{-}$) polarized components; $\omega_{c}$ is the central angular frequency of the continuous-wave control field. $\Delta_{4}$ ($\Delta_{3}$) is the one-photon  (two-photon) detuning; $\Delta_{2}$ is the Zeeman energy splitting induced by the applied magnetic field $B$ along $z$ direction.
(b)~Possible arrangement of experimental apparatus.
For more detail, see the text.
}
\label{Fig1}
\end{figure}
Levels $|1\rangle$, $|2\rangle$ and $|3\rangle$ are Zeeman-split sublevels of the atomic ground state, which are induced by a weak static magnetic field $B$ applied along the $z$ direction; level $|4\rangle$ is an excited state with spontaneous-emission decay rate denoted by $\Gamma_{4}$. The input probe field is linearly polarized, which can be regarded as a linear superposition of the right-circularly ($\sigma^{+}$) and the left-circularly ($\sigma^{-}$) polarized components, coupling to the transitions $|1\rangle\leftrightarrow|4\rangle$ and $|2\rangle\leftrightarrow|4\rangle$, respectively; the control field is $\pi$-polarized, coupling to the transition $|3\rangle\leftrightarrow|4\rangle$; see Fig.~\ref{Fig1}(a).
The probe (control) laser field is assumed to propagate along (perpendicularly to) $z$ direction~\cite{note01}; see Fig.~\ref{Fig1}(b).
$\Delta_{2}$ is the Zeeman energy splitting due to the externally applied magnetic field; $\Delta_{4}$ ($\Delta_{3}$) is one-photon (two-photon) detuning.
Note that the system consists of two EITs (called DEIT), which involve excitation channels $|1\rangle\leftrightarrow |4\rangle \leftrightarrow |3\rangle$ and $|2\rangle\leftrightarrow |4\rangle \leftrightarrow |3\rangle$,  respectively.

For simplicity, we assume the atomic gas is a cigar-shaped (with Fresnel number $F\approx 1$) or it is filled into
an optical waveguide with a small transverse size.
Therefore, the system can be approximately to be a one-dimensional one, as schematically shown in Fig.~\ref{Fig1}(b). The total electric field in the system reads
\begin{subequations}\label{Es0}
\begin{align}
& {\hat{\bf E}}(z,t)={\bf E}_{c}(z,t)+{\hat{\bf E}}_{p}(z,t), \\
& {\bf E}_{c}(z,t)={\bf e}_{c}{\cal E}_{c}e^{i(k_{c}z-\omega_{c}t)}+{\rm c.c.},\\
& {\hat{\bf E}}_{p}(z,t)={\hat{\bf E}}_{p1}(z,t)+{\hat{\bf E}}_{p2}(z,t),\\
& {\hat{\bf E}}_{pj}(z,t)={\bf e}_{pj}{\cal E}_{p0}{\hat E}_{pj}(z,t)e^{i(k_{p}z-\omega_{p}t)}+{\rm H.c.},
\end{align}
\end{subequations}
with $j=1,\,2$ and H.c. representing the complex conjugate. Here ${\bf e}_{c}$ and ${\cal E}_{c}$ are respectively the unit polarization vector and the control-field amplitude; ${\cal E}_{p0} \equiv \sqrt{\hbar\omega_{p}/(2\varepsilon_{0}V)}$ is the electric-field amplitude of a single probe photon, with ${\bf e}_{p1}=({\bf e}_{x}+i{\bf e}_{y})/\sqrt{2}$ [${\bf e}_{p2}=({\bf e}_{x}-i{\bf e}_{y})/\sqrt{2}$] the unit polarization vector of the $\sigma^{+}$ ($\sigma^{-}$)-polarized component; ${\hat E}_{p1}(z,t)$ [${\hat E}_{p2}(z,t)$] is the annihilation operator of probe photons in $\sigma^+$ ($\sigma^-$)-polarized component, obeying the commutation relation
\begin{eqnarray}
&& \left[{\hat E}_{pj}(z,t),{\hat E}_{pj'}^\dag(z',t)\right]=L\delta(z-z')\delta_{jj'},\,\,\,\, (j,j'=1,2),
\end{eqnarray}
where $L$ is the quantization length along the $z$-axis.

Under electric-dipole, rotating-waving, and paraxial approximations, the Hamiltonian of the system reads
\begin{align}\label{Hamiltonian}
\notag{\hat H}=
&-\frac{\hbar c}{L}\int_{-\infty}^{+\infty}dz\left[{\hat E}_{p1}^{\dag}\left(i\frac{\partial}{\partial z}\right){\hat E}_{p1}+{\hat E}_{p2}^{\dag}\left(i\frac{\partial}{\partial z}\right){\hat E}_{p2}\right] \\
\notag &-\frac{\hbar N}{L}\int_{-\infty}^{+\infty}dz\left(\sum_{\alpha=1}^4\Delta_{\alpha}
\hat{S}_{\alpha\alpha}+g_{p1}\hat{S}_{14}{\hat E}_{p1}\right.\\
&\hspace{2.0cm} \left.+g_{p2}\hat{S}_{24}{\hat E}_{p2}+\Omega_{c}\hat{S}_{34}+{\rm H.c.}\right).
\end{align}
Here, $N$ is total atomic number in the system; $\hat{S}_{\alpha\beta}(z,t)\equiv(\Delta N)^{-1}\sum_{l\in\Delta L} \hat{S}_{\alpha\beta}^{l}(t)\,(\alpha,\beta=1,2,3,4)$ are slowly-varying collective atomic transition operators, with the average being made over $\Delta N\,(\gg1)$ atoms within macroscopically small length $\Delta L$; the slowly-varying atomic operators related to the transition $|\alpha\rangle\leftrightarrow|\beta\rangle$ of $l$-th atom is defined by
\begin{equation}\label{Sjl}
\hat{S}_{\alpha\beta}^{l}(t)=\hat{\sigma}_{\beta\alpha}^{l}(t)
\exp\left\{i\left[(k_\beta-k_\alpha)z_{l}-(\omega_{\beta}+\Delta_{\beta}
-\omega_{\alpha}-\Delta_{\alpha}\right]t\right\},
\end{equation}
with $\hat{\sigma}_{\alpha\beta}^{l}\equiv|\alpha\rangle_{l\,l}\langle\beta|$.
$\hat{S}_{\alpha\beta}(z,t)$ obeys the commutation relation
\begin{eqnarray}\label{HSBGCR}
&& \left[\hat{S}_{\alpha\beta}(z,t),\hat{S}_{\alpha'\beta'}(z',t)\right]\notag\\
&& =\frac{L}{N}\delta(z-z')\left[\delta_{\alpha\beta'}\hat{S}_{\alpha'\beta}(z,t)
-\delta_{\alpha'\beta}\hat{S}_{\alpha\beta'}(z,t)\right];
\end{eqnarray}
$g_{p1}=({\bf e}_{p1}\cdot{\bf p}_{41}){\cal E}_{p}/\hbar$ [$g_{p2}=({\bf e}_{p2}\cdot{\bf p}_{42}){\cal E}_{p}/\hbar$] is the single-photon half Rabi frequency of the $\sigma^{+}$ ($\sigma^{-}$)-polarized component;
$\Omega_{c}\equiv ({\bf e}_{c}\cdot{\bf p}_{43}){\cal E}_{c}/\hbar$ is the half-Rabi frequency of the control field, with ${\bf p}_{\alpha\beta}$ the electric-dipole matrix element associated with the transition $|\alpha\rangle\leftrightarrow|\beta\rangle$.
The applied static magnetic field $B$ contributes to a Zeeman level shift $\Delta E_{B}=\mu_{B}g_{F}^{\alpha} m_{F}^{\alpha} B$,
with $\mu_{B}$, $g_{F}^{\alpha}$ and $m_{F}^{\alpha}$ being Bohr magneton, gyromagnetic factor, and magnetic quantum number of the atomic state $|\alpha\rangle$, respectively.
The detunings are given by $\Delta_{2}=-\mu_{21}B$, $\Delta_{3}=\omega_{p}-\omega_{c}-(\omega_{3}-\omega_{1})-\mu_{31}B$, and $\Delta_{4}=\omega_{p}-(\omega_{4}-\omega_{1})-\mu_{41}B$, with $\hbar\omega_{\alpha}$ being the eigenenergy of the state $|\alpha\rangle$ and $\mu_{\alpha\beta}=\mu_{B}(g_{F}^{\alpha} m_{F}^{\alpha}-g_{F}^{\beta} m_{F}^{\beta})/\hbar$.

The dynamics of the probe pulse and the atoms is controlled by MHL equations:
\begin{subequations}\label{HLM}
\begin{align}
& i\left(\frac{\partial}{\partial z}+\frac{1}{c}\frac{\partial}{\partial t}\right){\hat E}_{pj}+\frac{g_{pj}^\ast N}{c}\hat{S}_{4j}=0,\,\,\,\,(j=1,\,2)
\label{HLM0}\\
& i\frac{\partial}{\partial t}\hat{S}_{\alpha\beta}=\left[\hat{S}_{\alpha\beta},\frac{\hat{H}}{\hbar}\right]+i{\hat{\cal L}}(\hat{S}_{\alpha\beta})+i{\hat F}_{\alpha\beta}.\label{HLM1}
\end{align}
\end{subequations}
Here the term $\hat{\cal L}(\hat{S}_{\alpha\beta})$
is contributed by the dissipation process (described by spontaneous-emission rates $\Gamma_{\alpha\beta}$ and dephasing rates $\gamma_{\alpha\beta}^{\rm dep}$) in the system, and ${\hat F}_{\alpha\beta}$ are $\delta$-correlated Langevin noise operators describing the fluctuations associated with dissipations~\cite{Scully1997}. Explicit expression of the Heisenberg-Langevin equation [i.e. Eq.~(\ref{HLM}b)] is given in Appendix~\ref{app1}.

The proposed physical model described above can be easily realized by realistic experiments. One of candidates is the laser-cooled alkali $^{87}$Rb atomic gas. The atomic levels shown in Fig.~\ref{Fig1} can be chosen to be $|1\rangle=|5^2S_{1/2}, g_{F}^{1}=-1/2, m_{F}^{1} = -1 \rangle$,
$|2\rangle=|5^2S_{1/2}, g_{F}^{2}=-1/2, m_{F}^{2} = 1\rangle$, $|3\rangle=|5^2S_{1/2}, g_{F}^{3}=1/2, m_{F}^{3}=0\rangle$, and $|4\rangle = |5^2P_{1/2}, g_{F}^{4}=-1/6, m_{F}^{4}=0\rangle$. The other system parameters are given by $\Gamma_{4}= 2\pi\times5.75$ MHz, $\gamma_{31}^{\rm dep}=\gamma_{32}^{\rm dep}\approx2\pi\times1\,{\rm kHz}$~\cite{Steck}.

The model described above is similar to that employed in Ref.~~\cite{Zhang2021},
where the probe field is considered to be at single-photon level.
Differently, here we assume the probe field is a weakly nonlinear pulse, and hence it contains many photons.
Notice that, if the atoms are set to be in an exact two-photon resonance (i.e. the two-photon detuning $\Delta_3=0$), the system works in a regime of strict DEIT. In such a situation, the third-order Kerr nonlinearity of the system has a vanishing real part and a small imaginary part, which makes the existence of VOS and quantum squeezing impossible.
In the following, we shall assume that $\Delta_3$ takes a non-zero but small value, which makes the system work in a regime of perturbed DEIT.
Based on such a consideration, the quantum noise due to the spontaneous emission and dephasing can be largely suppressed, and the system can support very large real Kerr nonlinearity, and hence the formation of VOS and quantum squeezing of the probe field and the atomic spins (see below).
In addition, the one-photon detuning $\Delta_4$ is chosen to be larger than $\Gamma_{4}$, which makes the system work in a dispersive nonlinear regime. Thereby, the absorption of the VOS is negligibly small during propagation.

\subsection{Coupled quantum NLS equations for the propagation of the probe pulse}

Based on the basic spirit indicated above, we expect that an input probe pulse with the form of plane wave will be modulated due to the weak nonlinearity and dispersion in the system. Hence one can investigate the nonlinear dynamics of the system by adopting the method of amplitude equations, widely used in nonlinear wave theory~\cite{Jeffrey1982,Newell1992,Agrawal2001,Kivshar2003,Drummond2014}. By making a perturbation expansion on the MHL equations (\ref{HLM}) [generalizing the technique used in Ref.~\cite{Zhu2021} to the present two-component case], we can derive the following coupled quantum NLS equations for the envelope operators of the two polarization components of the probe pulse, given by
\begin{align}\label{CQNLS}
&\left[i\left(\frac{\partial}{\partial z}+\frac{1}{V_{gj}}\frac{\partial}{\partial t}\right)+{\rm Im}(K_{0j})\right]\hat{E}_{pj}-\frac{K_{2j}}{2}\frac{\partial^2}{\partial t^2}\hat{E}_{pj}\notag\\
&+|g_{p}|^{2}\left(W_{jj}\hat{E}_{pj}^{\dag}\hat{E}_{pj}+W_{j3-j}
\hat{E}_{p3-j}^{\dagger}\hat{E}_{p3-j}\right)\hat{E}_{pj}=i{\hat{\cal F}}_{pj}
\end{align}
($j=1,2$), when exact to cubic nonlinearity and second-order dispersion of the system. Here, $K_{0j}= K_{j}(\omega)|_{\omega=0}$, $V_{gj}^{-1}= K_{1j}\equiv (\partial K_{j}(\omega)/\partial\omega)|_{\omega=0}$ is the group velocity of the $j$-th component of the probe pulse, $K_{2j}\equiv(\partial^2K_{j}(\omega)/\partial\omega^2)|_{\omega=0}$ is the coefficient describing second-order dispersion (i.e. group velocity dispersion), with $K_{j}(\omega)$ being the linear dispersion relation of the $j$-th polarization component. Due to the symmetric configuration between the two EITs [see Fig.~\ref{Fig1}(a)], we have $g_{p2}\approx g_{p1}\equiv g_{p}$. In addition, $W_{jj}$ ($j=1,2$) are coefficients of self-phase modulation and $W_{j3-j}$ are coefficients of cross-phase modulation, which are proportional to the third-order nonlinear optical susceptibilities $\chi_{jl}^{(3)}$, i.e.
\begin{subequations}\label{chi3}
\begin{eqnarray}
&& \chi^{(3)}_{jj}=\frac{2c|{\bf e}_{pj}\cdot\mathbf{p}_{4j}|^{2}}{\hbar^2\omega_{p}}W_{jj},\,\,\,\,(j=1,2)\\
&& \chi^{(3)}_{jl}=\frac{2c|{\bf e}_{pj}\cdot\mathbf{p}_{3j}|^{2}}{\hbar^2\omega_{p}}W_{jl}, \,\,\,\,(j,l=1,2; j\neq l).
\end{eqnarray}
\end{subequations}
${\hat{\cal F}}_{pj}(z,t)$ ($j=1,2$) are $\delta$-correlated induced Langevin noise operators, which are necessary to make $\hat{S}_{jl}$ satisfy the Heisenberg commutation relations (\ref{HSBGCR}). The detailed derivation of Eqs.~(\ref{CQNLS}) and explicit expressions of $K_{j}(\omega)$, $W_{jj}$, $W_{jl}$ and ${\hat{\cal F}}_{pj}(z,t)$ are presented in Appendix~\ref{app2}.

Under condition $|\Omega_c|^2\gg \gamma_{4j}\gamma_{3j}$ ($j=1,2$) and large one-photon detuning $\Delta_4$, the loss of the probe pulse during propagation is very small, the Langevin noise plays no significant role and hence can be neglected (see the detailed explanation given at the end of the Appendix~\ref{app2}; similar discussions can also be found in
Refs.~\cite{Zhang2021,Gorshkov2007PRA1,Gorshkov2007PRA2}).
For the convenience of the following calculations, we write Eqs.~(\ref{CQNLS}) into the dimensionless form
\begin{align}\label{dml}
&i\left(\frac{\partial}{\partial s}+2\alpha_{j}\right)\hat{U}_{j}+ig_{\delta}\frac{\partial \hat{U}_{1}}{\partial \tau}+g_{Dj}\frac{\partial^2}{\partial \tau^2}\hat{U}_{j}\notag\\
&+2\left(g_{jj}\hat{U}_{j}^{\dag}\hat{U}_{j}+g_{j3-j}
\hat{U}_{3-j}^{\dagger}\hat{U}_{3-j}\right)\hat{U}_{j}=0
\end{align}
after neglecting the Langevin noise terms.
Here,  we have defined the dimensionless quantities $\hat{U}_{j}=\hat{E}_{pj}/\sqrt{n_{0}}$, $s=z/(2L_{D})$, $\tau=(t-z/V_{g})/t_{0}$,
$\delta=(1/V_{g1}-1/V_{g2})/2$, $V_g=2V_{g1}V_{g2}/(V_{g1}+V_{g2})$,
$\alpha_j=L_{D}/L_{j, A}$, $g_{\delta}={\rm sgn}(\delta)L_{D}/L_{\delta}$,
$g_{D1}=K_{21}/|K_{22}|$, $g_{D2}={\rm sgn}(-K_{22}\cdot W_{22})$, and $g_{jj(jl)}=W_{jj(jl)}/W_{22}$. In these definitions, $n_{0}$ ($\gg 1$) is the typical mean photon number in the probe pulse; $L_{\delta}=\tau_{0}/|\delta|$ is the typical group velocity mismatch
length; $t_{0}$, $L_{D}\equiv t_{0}^2/|K_{21}|$, and $L_{A,j}\equiv 1/{\rm Im}(K_{0j})$ are respectively typical time duration, dispersion length, and absorption length of the probe pulse. Since our main aim is to study the quantum squeezing of VOS, in the dimensionless NLS equations (\ref{dml}) we have assumed $L_{D}$ is equal to the typical nonlinear length of the system, which is defined by $L_{\rm NL}=[n_{0}|g_{p}|^2|W_{22}|]^{-1}$.

In general, $\alpha_1$ and $\alpha_2$ are not small and
the other coefficients in Eqs.~(\ref{dml}) are complex, which may result in significant loss of the probe pulse during propagation. However, under the conditions of the perturbed DEIT and large single-photon detuning $\Delta_4$, such a loss can be greatly suppressed.
This point can be clearly seen by using realistic physical parameters, given by
$\Omega_{c}=2.2\times10^{8}~ {\rm Hz}$, $\Delta_{2}=2\pi\times10^{3}~ {\rm Hz}$ (for $B=1 {\rm mG}$), $\Delta_{3}=2.91\times10^{6}~ {\rm Hz}$, $\Delta_{4}=2\times10^{8}~ {\rm Hz}$, and ${\cal N}_{a}\approx7.33\times10^{11} {\rm cm}^{-3}$ (atomic density). Then we can obtain the values of the coefficients of the second-order dispersion and the self- and cross-phase modulations, given in Table~\ref{TAB1}.
\begin{table}
\renewcommand\tabcolsep{17pt}
\centering
\caption{
\footnotesize {Coefficients $K_{jl}$ ($j,l=1,2$) of the second-order dispersion and the coefficients $W_{jl}$ ($j,l=1,2$) of the self- and cross-phase modulation, for system parameters taking to be
$\Omega_{c}=2.2\times10^{8}~ {\rm Hz}$, $\Delta_{2}=2\pi\times10^{3}~ {\rm Hz}$ (for $B=1 {\rm mG}$), $\Delta_{3}=2.91\times10^{6}~ {\rm Hz}$, $\Delta_{4}=2\times10^{8}~ {\rm Hz}$, and ${\cal N}_{a}\approx7.33\times10^{11} {\rm cm}^{-3}$.}
}
\vspace{0.2cm}
\label{TAB1}
\begin{tabular}{cc}
\hline\hline\vspace{-0.25cm}&\\
Parameters &
Value$\,({\rm cm^{-1}s^{2}})$
\vspace{0.1cm}
\\
\hline
&\vspace{-0.15cm}\\
$$
$K_{11}$&$(2.117+0.0047i)\times10^{-7}$
$$\\
$$
$K_{12}$&$(2.116+0.0047i)\times10^{-7}$
$$\\
$$
$K_{21}$&$(1.821+0.17i)\times10^{-15}$
$$\\
$$
$K_{22}$&$(1.820+0.16i)\times10^{-15}$
$$\\
$$
$W_{11}$&$-(2.521+0.00143i)\times10^{-17}$
$$\\
$$
$W_{22}$&$-(2.505+0.00139i)\times10^{-17}$
$$\\
$$
$W_{12}$&$-(2.507+0.0464i)\times10^{-17}$
$$\\
$$
$W_{21}$&$-(2.502+0.0461i)\times10^{-17}$
\vspace{0.1cm}
$$\\
\hline\hline
\end{tabular}
\end{table}
By setting $t_{0}=4.3\times10^{-8}$\,s,
we obtain
$L_{D}=L_{NL}\approx1.02~{\rm cm}$,
$\alpha_1\approx \alpha_2=1.3\times10^{-3}$,
$g_{\delta}=1.89\times10^{-4}$,
$g_{D1}=1.002+1\times10^{-6}i\approx1$, $g_{D2}=1$,
$g_{11}=1.0064+1\times10^{-5}i\approx 1$,
$g_{22}=1$,
$g_{12}=1.003+0.02i\approx 1$,
$g_{21}=0.995+0.02i\approx 1$.
We see that, indeed, $\alpha_1$ and $\alpha_2$ are very small and the imaginary parts of all other coefficients are much smaller than their corresponding real parts. This means that the loss of the probe pulse is small and can be taken as a small perturbation.

Based on such consideration, Eqs.~(\ref{dml}) can be simplified into the perturbed
quantum Manakov equations
\begin{align}\label{Simp-dml}
i\frac{\partial}{\partial s}\hat{U}_{j}+\frac{\partial^2}{\partial \tau^2}\hat{U}_{j}+2\left(\sum_{l=1,2}\hat{U}_{l}^{\dag}\hat{{ U}}_{l}\right)\hat{U}_{j}=R_j(\hat{U}_{1},\hat{U}_{2}),
\end{align}
where $R_j(\hat{U}_{1},\hat{U}_{2})$\,  ($j=1,2$) are perturbations.
The derivation of Eqs.~(\ref{Simp-dml}) from Eqs.~(\ref{dml}) is given in Appendix~\ref{app3}.
In the following calculations, we assume the system works in the regime of the perturbed  DEIT with a larger $\Delta_4$. For a short propagation distance (in the order of centimeter), the perturbations $R_j$ play a negligible role and can be neglected safely.

\begin{figure}
\centering
\includegraphics[width=0.95\columnwidth]{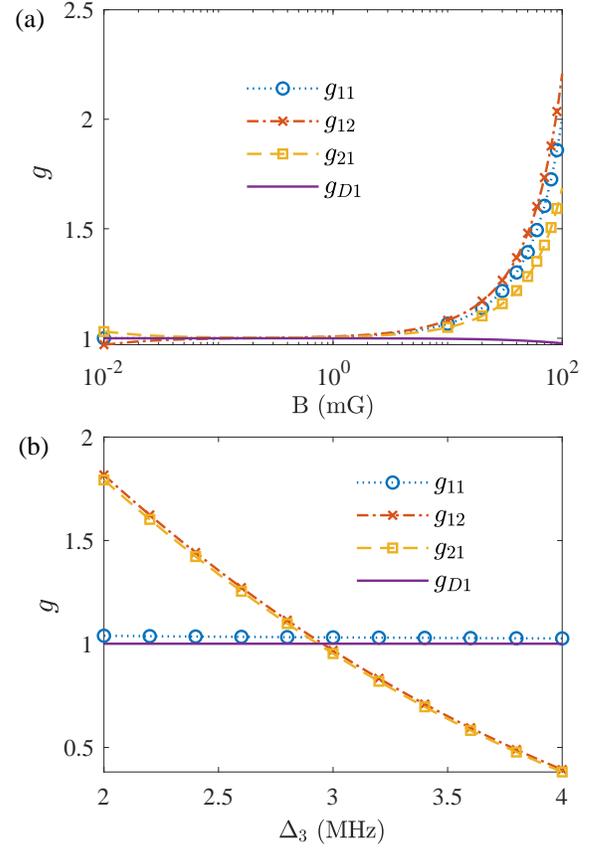}
\caption{
(a)~Dimensionless nonlinear coefficients $g_{jl}$ ($j,l=1,2$) and the ratio of the second-order dispersion $g_{D1}$ as functions of the magnetic field $B$ for the two-photon detuning $\Delta_{3}=2.9\times10^{6}~ {\rm Hz}$.
(b)~$g_{jl}$ ($j,l=1,2$) and $g_{D1}$ as functions of $\Delta_{3}$ for $B=5~{\rm mG}$.
}
\label{Fig2}
\end{figure}

In fact, the existence of non-zero perturbation terms $R_j(\hat{U}_{1},\hat{U}_{2})$ is mainly due to the existence of the  magnetic field $B$ and the two-photon detuning $\Delta_3$. Shown in Fig.~\ref{Fig2}
are  $g_{jl}$  ($j,l=1,2$) and $g_{D1}$   as functions of $B$ [panels (a)]  and $\Delta_3$  [panel (b)]. The other system parameters used for plotting the figure are the same as given above. We see that for small $B$
and for $\Delta_3\approx 2.9\times10^{6}~{\rm Hz}$, we have $g_{jl}\approx g_{D1}=1$.
In this situation, $R_j(\hat{U}_{1},\hat{U}_{2})$ can be indeed taken as perturbations.

With the above system parameters, we can estimate values of the self- and cross-Kerr nonlinear optical susceptibilities of the system based on the formula
(\ref{chi3}), given by
\begin{align}\label{chi31}
& \chi_{11}^{(3)}\approx\chi_{22}^{(3)}\approx\chi_{12}^{(3)}
\approx\chi_{21}^{(3)}=-3.69\times10^{-11}~{\rm m^{2}V^{-2}},
\end{align}
which are ten orders of magnitude larger than that obtained in comparison with conventional optical media (such as optical fibers). Physically, such giant Kerr nonlinearities are contributed by the nearly-resonant character and the DEIT in the system.

\section{Bi-orthonormal and complete eigenmodes of quantum fluctuations}\label{Sec3}

\subsection{Vector optical solitons}

In order to investigate the quantum fluctuations of VOS, we first consider the classical limit of the system. In such a situation the envelope operators $\hat{U}_{j}$ can be replaced by c-number envelope functions $V_{j}$.  Then Eqs.~(\ref{dml}) are reduced to
$i\left(\partial/\partial s+2\alpha_{j}\right)V_{j}+ig_{\delta}\partial V_{j}/\partial \tau+g_{Dj}\partial^2 V_{j}/\partial \tau^2
+2\left(g_{jj}|V_{j}|^{2}+g_{j3-j}|V_{3-j}|^{2}\right)V_{j}=0$, ($j=1,2$)
which are the classical two-component nonlinear NLS equations. In the past decades, there exist a large amount of works paid to the study on
the soliton-pair solutions supported by such equations~\cite{Newell1992,Agrawal2001,Kivshar2003,Lakoba1997,Skryabin2004,Assanto2008,
Guo2019,Ma2019}. Soliton-pair solutions in EIT systems have also been considered in recent years~\cite{Hang2008,Qi2011,Si2009,Chen2014}.

When neglecting the perturbation terms $R_j$, Eqs.~(\ref{Simp-dml}) is reduced into quantum Manakov equations
\begin{align}\label{QManakoveqs}
i\frac{\partial}{\partial s}\hat{U}_{j}+\frac{\partial^2}{\partial \tau^2}\hat{U}_{j}+2\left(\sum_{l=1,2}\hat{U}_{l}^{\dag}\hat{{ U}}_{l}\right)\hat{U}_{j}=0,
\end{align}

In classical limit, Eqs.~(\ref{QManakoveqs}) become classical Manakov equations, which admit the VOS solution~\cite{Newell1992,Agrawal2001,Kivshar2003}
\begin{align}\label{CSoliton0}
|V_0\rangle \equiv \binom{V_{1}}{V_{2}}=A_{0}e^{i\Theta_{0}}\binom{\cos\vartheta}{\sin\vartheta}
\,{\rm sech}\sigma,
\end{align}
with $\sigma=A_{0}(\tau-\tau_{0}-2p_{0}s)$,  $\Theta_{0}=p_{0} (\tau-\tau_{0})+(A_{0}^{2}-p_{0}^2)s+\theta_{0}$. Here $A_{0}$, $p_{0}$, $\tau_{0}$, and $\theta_{0}$ are four real parameters, determining the amplitude (or width), propagating velocity (or momentum), initial temporal position, and initial phase of each component of the VOS, respectively; the real parameter $\vartheta$ represents the ratio between the amplitudes of the two polarization components.

Using the system parameters given above, we obtain $V_{g2}\approx V_{g1}\equiv V_g=1.575\times10^{-4} c$. The propagating velocity $V_{\rm vos}$ of the above VOS solution has a small modification from $V_{g}$. For instance, for $p_0=1$
\begin{align}\label{Vs}
V_{\rm vos}=\frac{V_{g}}{1+\frac{V_{g}t_{0}}{L_{\rm disp}}p_{0}}\approx1.377\times10^{-4}  c,
\end{align}
which means that $V_{\rm vos}\approx V_{g}$, and both of $V_{g}$ and $V_{\rm vos}$ are much smaller than $c$ (the light speed in vacuum). This significant slowdown of the optical pulse is due to the DEIT effect induced by the control field.

The threshold power for creating the VOS can be estimated by calculating Poynting’s vector~\cite{Newell1992}. We obtain
\begin{align}\label{power}
{P}_{\max }=58.1~ \mu{\rm W}.
\end{align}
Thus very low input power is required for generation of the VOS, which is in contrast from the optical solitons generated in conventional optical media (e.g., optical fibers)~\cite{Newell1992,Agrawal2001}.

\subsection{Non-Hermitian four-component eigenvalue problem for quantum fluctuations}

Now we turn to investigate the quantum correction of the VOS in the system.
Making the transformation $\hat{U}_{j}=\hat{{\cal U}}_{j}\exp (iA_{0}^{2}s)$, Eqs.~(\ref{QManakoveqs}) become
\begin{align}\label{cnlse}
i\frac{\partial}{\partial s}\hat{{\cal U}}_{j}+\frac{\partial^2}{\partial \tau^2}\hat{{\cal U}}_{j}-A_{0}^{2}\hat{{\cal U}}_{j}+2\left(\sum_{l=1,2}\hat{{\cal U}}_{l}^{\dag}\hat{{\cal U}}_{l}\right)\hat{{\cal U}}_{j}=0.
\end{align}
The effective Hamiltonian described by Eqs.~(\ref{cnlse}) reads
\begin{align}\label{Heff0}
\hat{H}_{\rm eff}
&=\sum_{j=1,2}\int_{-\infty}^{+\infty}\hat{{\cal U}}_{j}^{\dagger}(\tau,s)
\left[-\frac{\partial^2}{\partial\tau^2}+A_{0}^{2}\right.\notag\\
&\left.-\hat{{\cal U}}_{j}^{\dag}(\tau,s)\hat{{\cal U}}_{j}(\tau,s)\right]\hat{{\cal U}}_{j}(\tau,s)d\tau\notag\\
&-2\int_{-\infty}^{+\infty}\hat{{\cal U}}_{1}^{\dag}(\tau,s)\hat{{\cal U}}_{2}^{\dag}(\tau,s)\hat{{\cal U}}_{1}(\tau,s)\hat{{\cal U}}_{2}(\tau,s)d\tau,
\end{align}
by which Eqs.~(\ref{cnlse}) can be written as the Heisenberg equations of motion $i\partial \hat{{\cal U}}_{j}/\partial s=\left[\hat{{\cal U}}_{j}, \hat{H}_{\rm eff}\right]$.

We assume that the mean photon number $n_0$ in the probe pulse is large, so that the quantum fluctuations are weaker compared with the VOS. The dimensionless probe field can be expressed by the Bogoliubov decomposition
\begin{equation}\label{BD}
\binom{\hat{{\cal U}}_{1}}{\hat{{\cal U}}_{2}}=\binom{V_{1}}{V_{2}}+\binom{\hat{v}_{1}}{\hat{v}_{2}},
\end{equation}
where $(\hat{v}_{1},\hat{v}_{2})^T$ ($T$ means transpose)
is a vector operator denoting the quantum fluctuations on the VOS background $(V_{1},V_{2})^T$; $\hat{v}_{1}$ and $\hat{v}_{2}$ satisfy the commutation relations
$[\hat{v}_{j}(\tau,s),\hat{v}_{l}^{\dag}(\tau',s)]=\delta_{jl}\delta(\tau-\tau')$.
Substituting (\ref{BD}) into (\ref{Heff0}) and neglecting the high-order terms of $\hat{v}_{j}$, we obtain the quadratic bose Hamiltonian
\begin{subequations}\label{aquareH}
\begin{align}
\hat{H}_{\rm eff}&=H_{0}+\hat{H}_{2},\\
H_{0}&=\sum_{j=1,2}\int_{-\infty}^{+\infty}d\tau\left[V_{j}\left(-\frac{\partial^2}{\partial\tau^2}+A_{0}^{2}-V_{j}^{2}\right)V_{j}-2V_{1}^{2}V_{2}^{2}\right],\\
{\hat H}_{2}& =\sum_{j=1,2}\int_{-\infty}^{\infty} d\tau\left[\hat{v}_{j}^\dag\left(-\frac{\partial^2}{\partial\tau^2}+A_{0}^{2}-4V_{j}^2\right)\hat{v}_{j}\right.\notag\\
&\left.-V_{j}^2(\hat{v}_{j}\hat{v}_{j}+\hat{v}_{j}^\dag\hat{v}_{j}^\dag)\right]-\int_{-\infty}^{\infty} d\tau2V_{1}V_{2}\left(\hat{v}_{1}^{\dagger}\hat{v}_{2}^{\dagger}+\hat{v}_{1}\hat{v}_{2}\right.\notag\\
&\left.+\hat{v}_{1}^{\dagger}\hat{v}_{1}+\hat{v}_{1}^{\dagger}\hat{v}_{2}+\hat{v}_{2}^{\dagger}\hat{v}_{1}+\hat{v}_{2}^{\dagger}\hat{v}_{2}\right).
\end{align}
\end{subequations}

With the VOS solution (\ref{CSoliton0}), the Bogoliubov decomposition (\ref{BD}) becomes
\begin{equation}\label{BD1}
\binom{\hat{{\cal U}}_{1}}{\hat{{\cal U}}_{2}}=e^{i\Theta_{0}'}\hat{T}\left(|\hat{V}_{0}\rangle+
|\hat{V}_{1}\rangle\right),
\end{equation}
with
\begin{subequations}
\begin{eqnarray}
&& |\hat{V}_{0}\rangle=A_{0}{\rm sech}[A_0(\tau-\tau_0-2p_0 s)]\binom{1}{0},\\
&& |\hat{V}_{1}\rangle=\binom{\hat{u}_{1}}{\hat{u}_{2}},
\end{eqnarray}
\end{subequations}
where $\hat{T}=\left(\begin{array}{cc}\cos\vartheta & -\sin\vartheta \\\sin\vartheta &\cos\vartheta\end{array}\right)$ is a rotation matrix, and $\Theta_{0}'=p_{0} (\tau-\tau_{0})-p_{0}^{2}s+\theta_{0}$.
Based on the expression (\ref{BD1}), the Heisenberg equations of motion for  $\hat{u}_{j}$ and $\hat{u}_{j}^{\dag}$ can be written as the form
\begin{align}\label{dyeq}
i\frac{\partial}{\partial s}|\hat{W}\rangle+A_{0}^{2}\,\hat{L}\,|\hat{W}\rangle=0
\end{align}
Here $|\hat{W}\rangle=(\hat{w}_{1},\hat{w}_{1}^{\dagger},\hat{w}_{2}, \hat{w}_{2}^{\dagger})^T$  is a vector operator of four components, with $\hat{w}_{j}\equiv \hat{u}_{j}/\sqrt{A_{0}}$  satisfying
the commutation relations
\begin{equation}
\left[\hat{w}_{j}(s,\sigma),\hat{w}_{l}^\dag(s,\sigma')\right]
=\delta_{jl}\delta(\sigma-\sigma').
\end{equation}
Here the definition of $\sigma$ is given in the classical VOS solution (\ref{CSoliton0}). In Eq.~(\ref{dyeq}),
$\hat{L}$ is a block-diagonal $4\times 4$ matrix describing the dynamics of quantum fluctuations on the VOS background, defined by
\begin{align}\label{L}
\hat{L}=\hat{L}_{a}\oplus\hat{L}_{b}
=\left(\begin{array}{cc}
\hat{L}_{a} & 0 \\
0 & \hat{L}_{b}
\end{array}\right)=\left(
\begin{array}{cccc}
{\cal M}  & {\cal N} & 0  & 0\\
         -{\cal N}    & -{\cal M} & 0  & 0\\
  0  & 0     & {\cal R}  & 0 \\
   0  & 0      & 0    & -{\cal R}
\end{array}\right),
\end{align}
with
$\hat{L}_{a}
\equiv\left(\begin{array}{cc}
{\cal M}  & {\cal N} \\
-{\cal N}    & -{\cal M}
\end{array}\right),
\hat{L}_{b}
\equiv\left(\begin{array}{cc}
{\cal R}  & 0 \\
0    & -{\cal R}
\end{array}\right)$,
${\cal M}\equiv \partial^{2}/\partial\sigma^{2}+4{\rm sech}^{2}\sigma-1$, ${\cal N}\equiv 2{\rm sech}^{2}\sigma$, and ${\cal R}\equiv\partial^{2}/\partial\sigma^{2}+2{\rm sech}^{2}\sigma-1$.

In order to consider all possible quantum fluctuations, one must solve the dynamical operator equation (\ref{dyeq}) and seek all possible eigenmodes of $\hat{ L}$. We require that these eigenmodes can constitute an eigenmode set that are orthogonal and complete, so that an arbitrary quantum fluctuation of the system can be expressed (expanded) by these eigenmodes. The ease of success for obtaining such an eigenmode set depends on the property of $\hat{L}$. On the one hand, from (\ref{L}) we see that $\hat{L}$ is a direct sum of $\hat{ L}_a$ and $\hat{ L}_b$. On the other hand, $\hat{L}_b$ is Hermitian; however, $\hat{L}_a$ is non-Hermitian but {\it pseudo-Hermitian}, with the adjoint operator  given by
$\hat{L}_a^{\dag}=\sigma_3\hat{L}_a\sigma_3=\left(
\begin{matrix}
{\cal M} & -{\cal N} \\
{\cal N} & -{\cal M}
\end{matrix}\right)$,
where $\sigma_3=\left(\begin{matrix}1 & 0\\0& -1\end{matrix}\right)$
(Pauli matrix). Thus, as a whole, $\hat{L}$ has the property
\begin{equation}
\hat{L}^{\dag}=\hat{\sigma}_3\hat{L}\hat{\sigma}_3,
\end{equation}
where $\hat{\sigma}_3\equiv\left(\begin{matrix}\sigma_3 & 0\\0& \sigma_3\end{matrix}\right)$. This fact tells us that
although $\hat{L}$ is not Hermitian, i.e. $\hat{L}^{\dagger}\neq \hat{L}$, but it is {\it pseudo-Hermitian}. In recent years, it has been proved that
pseudo-Hermitian operators possess all-real spectra (eigenvalues). If one is able to find the all eigenmodes of $\hat{L}$ and $\hat{L}^{\dag}$, complete and bi-orthonormal eigenemode bases can be constructed in mutually dual function spaces of $\hat{L}$ and $\hat{ L}^{\dag}$~\cite{Vladimir2016,Ashida2020}, by which the effective Hamiltonian (\ref{aquareH}) can be diagonalized.

To this end, we make the Bogoliubov transformation~\cite{Zhu2021,Zhu2022}
\begin{subequations}\label{w0}
\begin{align}
& {\hat w}_{1}(\sigma,s)
=\sum_{f}\left[u_{f}^{(a)}(\sigma)\hat{a}^{(a)}_{f}(s)+v^{(a)\ast}_{f}(\sigma)
\hat{a}_{f}^{(a)\dagger}(s)\right],\\
& {\hat w}_{2}(\sigma,s)
=\sum_{f}\left[u_{f}^{(b)}(\sigma)\hat{a}^{(b)}_{f}(s)+v^{(b)\ast}_{f}(\sigma)
\hat{a}_{f}^{(b)\dagger}(s)\right].
\end{align}
\end{subequations}
Here $\hat{a}^{(\alpha)}_{f}$ and $\hat{a}^{(\alpha) \dag}_f$
are respectively annihilation and creation operators of photons
for the mode $f$, satisfying the commutation relations $[\hat{a}^{(\alpha)}_{f}(s),\hat{a}^{(\alpha')\dagger}_{f'}(s)]
=\delta_{\alpha \alpha'}\delta_{f f'}$;
$u^{(\alpha)}_{f}(\sigma)$ and $v^{(\alpha)}_{f}(\sigma)$
are mode functions  ($\alpha=a, b$).

Assuming $|\hat{W}(\sigma,s)\rangle=|\hat{W}(\sigma)\rangle \exp (iA_{0}^{2}\lambda s)$, substituting it into Eq.~(\ref{dyeq}), and using the
Bogoliubov transformation (\ref{w0}),
we obtain the eigenvalue equations (i.e. BdG equations) for the non-Hermitian operator $\hat{L}$:
\begin{align}\label{BdGE}
\hat{L}\,
|\Psi_f(\sigma)\rangle=\lambda_f|\Psi_f(\sigma)\rangle.
\end{align}
Here, $\lambda_f$ is eigenvalue;  $|\Psi_f(\sigma)\rangle=\left(
\begin{array}{cccc}
u^{(a)}_f, &v^{(a)}_f, &u^{(b)}_f, &v^{(b)}_f
\end{array}\right)^{T}$ is the corresponding eigenvector, which has four components. Due to the block-diagonal structure of $\hat{L}$, we can independently solve the eigenvalue equations for each of the two diagonal blocks in Eq.~(\ref{BdGE}), which read
\begin{subequations}
\begin{eqnarray}
&& \hat{L}_{a}|\Psi^{(a)}_f(\sigma)\rangle=\lambda^{(a)}_f|\Psi^{(a)}_f(\sigma)\rangle,\\
&& \hat{L}_{b}|\Psi^{(b)}_f(\sigma)\rangle=\lambda^{(b)}_f|\Psi^{(b)}_f(\sigma)\rangle,
\end{eqnarray}
\end{subequations}
where $\lambda^{(a)}_f$ and $\lambda^{(b)}_f$ are eigenvalues,
$|\Psi^{(a)}_f\rangle=\left(u^{(a)}_f, v^{(a)}_f\right)^T$ and $|\Psi^{(b)}_f\rangle=\left(u^{(b)}_f,v^{(b)}_f\right)^T$ are two-component eigenvectors of the operators $\hat{L}_{a}$ and $\hat{L}_{b}$ , respectively. The four-component vector $|\Psi(\sigma)_f\rangle$ can be generally written into the form
\begin{align}
|\Psi_f(\sigma)\rangle
=\binom{1}{0}\otimes |\Psi^{(a)}_f(\sigma)\rangle\,\,\,\, {\rm or} \,\,\binom{0}{1}\otimes |\Psi^{(b)}_f(\sigma)\rangle.
\end{align}

\subsection{Eigenmodes of the quantum fluctuation and their bi-orthonormality and completeness}

The eigenvalues and eigenmodes of the operator $\hat{L}_{a}$ and $\hat{L}_{a}^{\dag}$ are known in the study on the quantum perturbation theory of bright solitons with a single component~\cite{Zhu2021}, which include continuous and discrete spectra. The eigen-equations and eigenmodes of the continuous spectrum (with eigenvalues $\lambda_k^{(a)}=k^2+1$) read
\begin{subequations}
\begin{eqnarray}\label{Lacont}
&& \hat{L}_{a}|\Psi^{(a)}_k(\sigma)\rangle=(k^2+1)|\Psi^{(a)}(\sigma)\rangle,\,\,\,\,
(-\infty < k <\infty)\label{Lacont1}\\
&& |\Psi_{k}^{(a)}(\sigma)\rangle=\binom{u_{k}^{(a)}}{v_{k}^{(a)}}
=-\frac{{\rm e}^{ik\sigma}}{\sqrt{2\pi}(k+1)^{2}}\notag\\
&& \hspace{1.cm}\times\left[{\rm sech}^{2}\sigma\binom{1}{1}+\left(k^{2}+2ik\tanh\sigma-1\right)\binom{1}{0}\right].
\label{Lacont2}
\end{eqnarray}
\end{subequations}
The discrete spectrum of $\hat{L}_a$ contains two zero modes
(with the eigenvalue $\lambda^{(a)}_{1}=\lambda^{(a)}_{2}=0$).
The eigen-equations and eigen-functions are given by
\begin{subequations}\label{Ladis}
\begin{eqnarray}
&& \hat{L}_{a}|\Psi^{(a)}_{n}(\sigma)\rangle=0,\,\,\,\,(n=1,2)\label{Ladis1}\\
&& |\Psi^{(a)}_{1}(\sigma)\rangle=\binom{u_{1}^{(a)}}{v_{1}^{(a)}}
=\frac{{\rm sech}\sigma}{2}\binom{2-\sigma\tanh\sigma}{-\sigma\tanh\sigma},\\
&& |\Psi^{(a)}_{2}(\sigma)\rangle=\binom{u_{2}^{(a)}}{v_{2}^{(a)}}
=\frac{{\rm sech}\sigma}{2}\binom{\tanh\sigma+\sigma}{\tanh\sigma-\sigma}.\label{Ladis2}
\end{eqnarray}
\end{subequations}
The eigenvalues and the eigenmodes  of $\hat{L}_a^{\dag}$ are given by
\begin{subequations}\label{Ldacont}
\begin{eqnarray}
&& \hat{L}_{a}^{\dag}|\Phi^{(a)}_k(\sigma)\rangle=(k^2+1)|\Phi^{(a)}_k(\sigma)\rangle,
\label{Ldacont1}\\
&& |\Phi_{k}^{(a)}(\sigma)\rangle=\sigma_3 |\Psi_{k}^{(a)}(\sigma)\rangle.
\label{Ldacont2}
\end{eqnarray}
\end{subequations}
for the continuous spectrum, and
\begin{subequations}\label{Ldadis}
\begin{eqnarray}
&& \hat{L}_{a}^{\dag}|\Phi^{(a)}_n(\sigma)\rangle=0, \,\,\,\,(n=1,2)
\label{Ldadis1}\\
&& |\Phi_{n}^{(a)}(\sigma)\rangle=\sigma_3 |\Psi_{n}^{(a)}(\sigma)\rangle,
\label{Ldadis2}
\end{eqnarray}
\end{subequations}
for the discrete spectrum.

The two eigenmode sets of $\hat{L}_a$ and $\hat{L}_a^{\dag}$ given above constitute two mutually dual function spaces, with the bases respectively given by
$\{|\Psi_f\rangle;f=n,k\}$ and $\{ |\Phi_f\rangle;f=n,k\}$, which are bi-orthonormal and complete in the following sense:
\begin{subequations}
\begin{eqnarray}
&&\langle\Phi^{(a)}_{f}(\sigma)|\Psi^{(a)}_{f'}(\sigma)\rangle=\delta_{ff'},
\label{orthonormality1}\\
&&\sum_{n=1,2}|\Phi^{(a)}_{n}(\sigma)\rangle\langle
\Psi^{(a)}_{n}(\sigma')|
+\int_{-\infty}^{+\infty}dk|\Phi^{(a)}_k(\sigma)\rangle\langle\Psi^{(a)}_k(\sigma')| \notag\\
&& =I\delta(\sigma-\sigma'),\label{completeness1}
\end{eqnarray}
\end{subequations}
with $I$ is $2\times2$ identity matrix. Here, the scalar product between the right vectors
$\{|\Psi_{f}^{(a)}\rangle\}$ and the left vectors $\{\langle \Phi_{f'}^{(a)}|\}$ is defined by
\begin{align}\label{inner}
\langle\Phi_{f}^{(a)}|\Psi_{f'}^{(a)}\rangle=\int_{-\infty}^{\infty}d\sigma
\langle \Phi_{f}^{(a)} (\sigma)|\Psi_{f'}^{(a)}(\sigma)\rangle.
\end{align}

 It is easy to get the eigenvalues and eigenmodes of $\hat{L}_b$ because it is a Hermitian operator. $\hat{L}_b$ has two branches of continuous spectra, which satisfy the eigen equations
\begin{align}\label{Lbconteq}
& \hat{L}_b |\Psi_{k\pm}^{(b)}(\sigma)\rangle=\pm(k^{2}+1)|\Psi_{k\pm}^{(b)}(\sigma)\rangle
\end{align}
($-\infty < k <\infty$). The corresponding  eigenmodes are given by
\begin{subequations}\label{Lbcontmodes}
\begin{align}
& |\Psi_{k+}^{(b)}(\sigma)\rangle=\binom{u_{k+}^{(b)}}{v_{k+}^{(b)}}=\frac{{\rm e}^{ik\sigma}(-ik+\tanh\sigma)}{\sqrt{2\pi}(-ik+1)}\binom{0}{1},\\
& |\Psi_{k-}^{(b)}(\sigma)\rangle=\binom{u_{k-}^{(b)}}{v_{k-}^{(b)}}=\frac{{\rm e}^{ik\sigma}(-ik+\tanh\sigma)}{\sqrt{2\pi}(-ik+1)}\binom{1}{0}.
\end{align}
\end{subequations}
$\hat{L}_b$ also has two zero modes (i.e. $\lambda^{(b)}_{1}=\lambda^{(b)}_{2}=0$), satisfying
\begin{align}\label{Lbdisceq}
& \hat{L}_b |\Psi_{n}^{(b)}(\sigma)\rangle=0,\,\,\,\,(n=1,2)
\end{align}
with
\begin{subequations}\label{Lbdiscmode}
\begin{align}
& |\Psi_{1}^{(b)}(\sigma)\rangle=\binom{u_{1}^{(b)}}{v_{1}^{(b)}}=\frac{{\rm sech}\sigma}{\sqrt{2}}\binom{1}{-1},\\
& |\Psi_{2}^{(b)}(\sigma)=\binom{u_{2}^{(b)}}{v_{2}^{(b)}}=\frac{{\rm sech}\sigma}{\sqrt{2}}\binom{1}{1}.
\end{align}
\end{subequations}

The eigenmode set of $\hat{L}_b$ constitutes a Hilbert space, in which they fulfill the orthonormality and completeness relations:
\begin{subequations}
\begin{eqnarray}
&&\langle\Psi^{(b)}_{f}(\sigma)|\Psi^{(b)}_{f'}(\sigma)\rangle=\delta_{ff'},
\label{orthonormality2}\\
&&\sum_{n=1}^2|\Psi^{(b)}_{n}(\sigma)\rangle\langle\Psi^{(b)}_{n}(\sigma')|
+\int_{-\infty}^{+\infty}dk
\sum_{\alpha=+,-}|\Psi^{(b)}_{k\alpha}(\sigma)\rangle\langle\Psi^{(b)}_{k\alpha}(\sigma')| \notag\\
&& =I\delta(\sigma-\sigma'),\label{completeness2}
\end{eqnarray}
\end{subequations}

Based on the orthonormal and complete properties of the eigenmodes of $\hat{L}_a$ and $\hat{L}_b$ respectively, it is easy to prove the bi-orthonormality and completeness of the eigenmodes of $\hat{L}$  due to its block-diagonal structure. They are given by
\begin{subequations}
\begin{eqnarray}
&&\langle\Phi_{f}(\sigma)|\Psi_{f'}(\sigma)\rangle=\delta_{ff'},\,\,\,\,(f,f'=n,k; n=1,2)
\label{orthonormality1}\\
&&\sum_{n=1,2}|\Psi_{n}(\sigma)\rangle\langle\Phi_{n}(\sigma')|
+\int_{-\infty}^{+\infty}dk|\Psi_k(\sigma)\rangle\langle\Phi_k(\sigma')| \notag\\
&& =\hat{I}\delta(\sigma-\sigma').\label{completeness1}
\end{eqnarray}
\end{subequations}
Here $\hat{I}$ is $4\times4$ identity matrix, $|\Psi_f(\sigma)\rangle=\binom{1}{0}\otimes |\Psi^{(a)}_f(\sigma)\rangle+ \binom{0}{1}\otimes |\Psi^{(b)}_f(\sigma)\rangle$ and $\langle\Phi_f(\sigma)|=\binom{1}{0}\otimes \langle\Phi^{(a)}_f(\sigma)|+ \binom{0}{1}\otimes \langle\Psi^{(b)}_f(\sigma)|$.

\subsection{Diagonalization of the effective Hamiltonian}

Based on the results above, the Bogoliubov transformation (\ref{w0}) can be written into the form
\begin{subequations}\label{w1}
\begin{align}
{\hat w}_{1}(\sigma,s)
=&\sum_{n=1}^2\left[u_{n}^{(a)}(\sigma)\hat{a}^{(a)}_{n}(s)+v^{(a)\ast}_{n}(\sigma)
\hat{a}_{n}^{(a)\dagger}(s)\right]\notag\\
&+\int dk\,\left[u^{(a)}_{k}(\sigma)\hat{a}^{(a)}_{k}(s)+v^{(a)\ast}_{k}(\sigma)
\hat{a}^{(a)\dagger}_{k}(s)\right],\\
{\hat w}_{2}(\sigma,s)
=&\sum_{n=1}^2\left[u_{n}^{(b)}(\sigma)\hat{a}^{(b)}_{n}(s)+v^{(b)\ast}_{n}(\sigma)
\hat{a}_{n}^{(b)\dagger}(s)\right]\notag\\
&+\int dk\,\left[u^{(b)}_{k}(\sigma)\hat{a}^{(b)}_{k}(s)+v^{(b)\ast}_{k}(\sigma)
\hat{a}^{(b)\dagger}_{k}(s)\right],
\end{align}
\end{subequations}
Here the indices $n$ and $k$ are quantum numbers denoting respectively discrete and continuous modes;
$\hat{a}^{(\alpha)}_{n}(s)$ and $\hat{a}^{(\alpha)}_{k}(s)$  ($\alpha=a,b$) are respectively annihilation operators of photons for the discrete and continuous modes, satisfying respectively  the commutation relations $[\hat{a}^{(\alpha)}_{n}(s),\hat{a}^{(\alpha')\dagger}_{n'}(s)]
=\delta_{\alpha \alpha'}\delta_{nn'}$ and $[\hat{a}^{(\alpha)}_{k}(s),\hat{a}^{(\alpha')\dagger}_{k'}(s)]=\delta_{\alpha
\alpha'}\delta (k-k')$;
$u^{(\alpha)}_{n}(\sigma)$, $v^{(\alpha)}_{n}(\sigma)$, $u^{(\alpha)}_{k}(\sigma)$, and $v^{(\alpha)}_{k}(\sigma)$
are mode functions for the discrete and continuous spectra, respectively.

With these exact results, we can diagonalize the effective Hamiltonian (\ref{aquareH}) into the form
\begin{align}\label{diahamiltonian}
\hat{H}_{\rm eff}=\frac{4}{3}A_{0}^{3}+& A_{0}^{2}\left\{
\left[\hat{P}_{2}^{(a)}(s)\right]^2-\left[\hat{Q}_{1}^{(a)}(s)\right]^2\right.
\notag\\
&\hspace{4mm}+\left.\sum_{\alpha=a,b}\int_{-\infty}^{+\infty}dk\lambda_{k}^{(\alpha)}
\hat{a}_{k}^{(\alpha)\dagger}(s)\hat{a}_{k}^{(\alpha)}(s)\right\},
\end{align}
with ${\hat Q}_{n}^{(\alpha)}=\left(\hat{a}^{(\alpha)}_{n}+\hat{a}_{n}^{(\alpha)\dag}\right)
/\sqrt{2}$,
${\hat P}_{n}^{(\alpha)}=\left(\hat{a}^{(\alpha)}_{n}
-\hat{a}_{n}^{(\alpha)\dag}\right)/(\sqrt{2}i)$ being respectively
the ``position'' operators and ``momentum'' operators related to the discrete-spectrum eigenmodes, satisfying commutation relations $[{\hat Q}^{(\alpha)}_{n},{\hat P}^{(\alpha')}_{n'}]=i\delta_{\alpha \alpha'}\delta_{nn'}$\, ($\alpha, \alpha'=a,b$; $n, n'=1,2$). Terms related to
$\hat{P}_{2}^{(a)}$ and $\hat{Q}_{1}^{(a)}$ can be understood to be induced by the deformation of the VOS, while the term related to $\hat{a}_{k}^{(\alpha)}$ is due to the radiations from the VOS.

Note that the four zero modes obtained above are not Goldstone modes, though their origin have some similarities to that of Goldstone bosons. The occurrence of these zero modes is due to the existence of the VOS, which is inhomogeneous in space and time~\cite{Zhu2021,Zhu2022}. For detailed discussions on quantum solitons and zero modes, see Ref.~\cite{Faddeev1978,Lee1981,Raj1987}.

\section{Quantum squeezing of light polarizations and atomic spins}\label{Sec4}

\subsection{Quantum dynamics of the vector optical soliton}

Based on the diagonalized effective Hamiltonian (\ref{diahamiltonian}), we can obtain the dynamics equations describing the quantum fluctuations of the VOS:
\begin{subequations}\label{HE0}
\begin{align}
& \frac{\partial}{\partial s}\hat{Q}^{(a)}_{1}=0,\\
& \frac{\partial}{\partial s}\hat{P}^{(a)}_{1}-2A_{0}^{2}\hat{Q}^{(a)}_{1}=0, \label{xi}\\
&  \frac{\partial}{\partial s}\hat{P}^{(a)}_{2}=0,\\
& \frac{\partial}{\partial s}\hat{Q}^{(a)}_{2}-2A_{0}^{2}\hat{P}^{(a)}_{2}=0. \label{alpha},\\
& \frac{\partial}{\partial s}\hat{Q}^{(b)}_{n}=0,\\
& \frac{\partial}{\partial s}\hat{P}^{(b)}_{n}=0,\\
& i\frac{\partial}{\partial s}\hat{a}^{(\alpha)}_{k}-A_{0}^{2}\lambda^{(\alpha)}_{k}\hat{a}^{(\alpha)}_{k}=0
\,\,\,\, (\alpha=a,b).
\end{align}
\end{subequations}
It is easy to get the exact solutions of these equations, which are given by
\begin{subequations}\label{SolutionofHE}
\begin{align}
& \hat{Q}^{(a)}_{1}(s)=\hat{Q}^{(a)}_{1}(0), \label{SolutionofHE1}\\
& \hat{P}^{(a)}_{1}(s)=2A_{0}^{2}\hat{Q}^{(a)}_{1}(0)s+\hat{P}^{(a)}_{1}(0),
\label{SolutionofHE2}\\
& \hat{Q}^{(a)}_{2}(s)=2A_{0}^{2}\hat{P}^{(a)}_{2}(0)s+\hat{Q}^{(a)}_{2}(0),
\label{SolutionofHE3}\\
& \hat{P}^{(a)}_{2}(s)=\hat{P}^{(a)}_{2}(0),\label{SolutionofHE4}\\
& \hat{Q}^{(b)}_{n}(s)=\hat{Q}^{(b)}_{n}(0), \label{SolutionofHE5} \\
& \hat{P}^{(b)}_{n}(s)=\hat{P}^{(b)}_{n}(0),\,\,\, (n=1,2) \label{SolutionofHE6}\\
& \hat{a}^{(\alpha)}_{k}=\hat{a}^{(\alpha)}_{k}(0)\exp(-iA_{0}^{2}
\lambda^{(\alpha)}_{k}s),\,\,\,\, (\alpha=a,b) \label{SolutionofHE7}
\end{align}
\end{subequations}
where $\hat{Q}^{(\alpha)}_{n}(0)$, $\hat{P}^{(\alpha)}_{n}(0)$, $\hat{a}^{(\alpha)}_{k}(0)$ are the values of $\hat{Q}^{(\alpha)}_{n}(s)$, $\hat{P}^{(\alpha)}_{n}(s)$, $\hat{a}^{(\alpha)}_{k}(s)$ at $s=0$, respectively.
From these solutions we can obtain the following conclusions:
(i)~From (\ref{SolutionofHE1})-(\ref{SolutionofHE4}), we see that during propagation $\hat{Q}^{(a)}_{1}$  and  $\hat{P}^{(a)}_{2}$ remain unchanged; however, ${\hat P}^{(a)}_{1}$ and ${\hat Q}^{(a)}_{2}$ are changed and they become correlated with $\hat{Q}^{(a)}_{1}$  and  $\hat{P}^{(a)}_{2}$.
A direct outcome of such a correlation between $\hat{Q}^{(a)}_n$ and $\hat{P}^{(a)}_n$ ($n=1,2$) is the induction of the phase diffusion and position spreading of the VOS, which are contributed by the self- and cross-Kerr nonlinearities in the system.
(ii)~Interestingly, from (\ref{SolutionofHE5})-(\ref{SolutionofHE6}) we see that
the two zero modes of the Hermitian operator $\hat{L}_b$ have no effect on the VOS deformation. This property is the result of the
left-right
configuration symmetry between the two EITs in the system [see in Fig.~\ref{Fig1}(a)].
If this symmetry is broken (e.g. by using a larger magnetic field $B$, which is not considered here), the Hermitian property of $\hat{L}_b$ will be lost and hence its zero modes will contribute to the deformation of the VOS.
(iii)~The continuous modes of both $\hat{L}_a$ and $\hat{L}_b$  have contributions to quantum fluctuations. However, these (radiation) modes display only a simple effect, i.e. each of them contributes a constant phase shift to itself (which is also caused by the Kerr nonlinearities in the system).

\subsection{Polarization squeezing of the probe pulse}

Based on the results obtained above, we now investigate the quantum squeezing of the VOS. Because the quantum fluctuations from the continuous modes are much smaller comparing with those from the zero modes~\cite{Zhu2021,Zhu2022,HausJOSAB1990,YLai1993}, they will be neglected in the following calculations.

The quantum property of the probe field with the two polarization components can be described by the following Stokes operators~\cite{Corney2006,Corney2008}
\begin{subequations}\label{stokesope}
\begin{align}
& \hat{s}_0\equiv \hat{N}_{11}+\hat{N}_{22},\\
& \hat{s}_1=\hat{N}_{12}+\hat{N}_{21},\\
& \hat{s}_2=i\left(\hat{N}_{21}-\hat{N}_{12}\right),\\
& \hat{s}_3=\hat{N}_{11}-\hat{N}_{22},
\end{align}
\end{subequations}
where
$\hat{N}_{j j'}=\int d\tau \hat{E}_{pj}^{\dag} (s,\tau)\hat{E}_{p j'} (s,\tau)$, with $j,j'=1,2$. These operators are Hermitian (thus observables), and satisfy the commutation relations of angular momentum
\begin{align}\label{SU2cr}
\left[\hat{s}_{0},\hat{s}_{i}\right]=0,\,\hspace{0.5 cm}\,\left[\hat{s}_{i},\hat{s}_{j}\right]=2i\epsilon_{ijk}\hat{s}_{k},
\end{align}
where $i,j,k=1,2,3$ and $\epsilon_{ijk}$ is antisymmetric unit (Levi Civita) tensor. In these operators, $\hat{s}_0$ corresponds to the intensity of the probe pulse, $\hat{s}_3$ describes the number difference, and $\hat{s}_1$ and $\hat{s}_2$ describe the relative phase difference between the two components.
Due to their quantum property, the variances of the Stokes operators obey the Heisenberg uncertainty relation
\begin{align}\label{uncr}
\langle(\Delta\hat{s}_{i})^{2}\rangle\langle
(\Delta\hat{s}_{j})^{2}\rangle\geq\epsilon_{ijk}|\langle\hat{s}_{k}\rangle|^{2}.
\end{align}
Here $\langle(\Delta\hat{s}_{i})^2\rangle\equiv\langle(\hat{s}_{i}
-\bar{s}_{i})^2\rangle$ is the variance of the  Stokes operator $\hat{s}_{i}$, with $\bar{s}_{i}\equiv \langle \hat{s}_{i}\rangle$.

Although the uncertainty relation (\ref{uncr}) is state-dependent, one is always able to find pairs of maximally conjugate operators by defining a Stokes basis in which only one Stokes operator has a non-zero expectation value. This can be realized by considering a special polarization state for which $\langle\hat{s}_{i}\rangle=\langle\hat{s}_{j}\rangle=0$ and $\langle\hat{s}_{k}\rangle=\langle\hat{s}_{0}\rangle\neq 0$, where $\hat{s}_{i}$,
$\hat{s}_{i}$, and $\hat{s}_{k}$ are Stokes operators that are orthogonal each other~\cite{Corney2006,Corney2008}. In this basis, there is only one nontrivial uncertainty inequality,
given by
\begin{align}\label{uncr1}
\langle(\Delta\hat{s}_{i})^{2}\rangle\langle
(\Delta\hat{s}_{j})^{2}\rangle\geq|\langle\hat{s}_{k}\rangle|^{2}
=|\langle\hat{s}_{0}\rangle|^{2}.
\end{align}
The polarization squeezing is achieved if
\begin{align}
\langle(\Delta \hat{s}_{i})^{2}\rangle<|\langle\hat{s}_{0}\rangle|<
\langle(\Delta \hat{s}_{j})^{2}\rangle.
\end{align}

The choice of the conjugate operator pair ($\hat{s}_i$, $\hat{s}_j$) in (\ref{uncr1}) is not unique. In fact,
an infinite set of such operator pairs exist in the plane of  $\hat{s}_i$-$\hat{s}_j$. Since $\langle\hat{s}_i\rangle$ and $\langle\hat{s}_j\rangle$ have zero mean value, this plane is called dark plane.
The direction of $\langle\hat{\bf s}\rangle$ defines a axis (called Stokes axis), with $|\langle \hat{\bf s}\rangle |=|\langle \hat{s}_0\rangle |$ along this axis.
Generally, a dark-plane operator can be defined as
\begin{align}\label{sstokes0}
\hat{s}_{\theta}=\cos\theta\hat{s}_{i}+\sin\theta\hat{s}_{j},
\end{align}
with $\theta$ being the detection angle in the dark plane relative to $\hat{s}_{i}$.  Thus the polarization squeezing occurs when
\begin{eqnarray}
\langle(\Delta\hat{s}_{\theta})^{2}\rangle<|\langle \hat{s}_0\rangle|<\langle(\Delta\hat{s}_{\theta+\pi/2})^{2}\rangle.
\end{eqnarray}
Usually, the squeezing ratio defined by $R_{p}(\theta)= \langle(\Delta\hat{s}_{\theta})^{2}\rangle/|\langle \hat{s}_0\rangle|$
is taken to estimate the amount (or degree) of the polarization squeezing.

In the above expressions, the quantum average of operator $\hat{O}$ is defined by $\langle\hat{O}\rangle \equiv\langle\Psi_0|\hat{O}|\Psi_0\rangle$.  Here we assume  $|\Psi_0\rangle=|n_{0},0,0,\cdots\rangle$ is the input quantum state of the probe pulse; it is a coherent state, with the photon number $n_0>>1$ and $n_f^{(a)}=n_f^{(b)}=0$ (where $n_f^{(a)}$ and $n_f^{(b)}$ are occupation numbers of the modes of $\hat{L}_a$ and $\hat{L}_b$, respectively). In the following calculations, we take $n_0=1.55\times 10^4$.

Because $\hat{E}_{pj}=\sqrt{n_0}\hat{U}_{j}e^{iA_0^2 s}$, with $\hat{U}_{j}\equiv V_{j}+\hat{u}_{j}$, we have $\hat{N}_{j j'}=n_0\int d\tau \hat{U}_{j}^{\dag} (s,\tau)\hat{U}_{j'} (s,\tau)$. For convenience, we define the reduced Stokes operators
\begin{subequations}\label{stokesope1}
\begin{align}
& \hat{\cal S}_0\equiv \hat{\cal N}_{11}+\hat{\cal N}_{22},\\
& \hat{\cal S}_1=\hat{\cal N}_{12}+\hat{\cal N}_{21},\\
& \hat{\cal S}_2=i\left(\hat{\cal N}_{21}-\hat{\cal N}_{12}\right),\\
& \hat{\cal S}_3=\hat{\cal N}_{11}-\hat{\cal N}_{22},
\end{align}
\end{subequations}
where
$\hat{\cal N}_{j j'}=\int d\tau \hat{U}_{j}^{\dag} (s,\tau)\hat{U}_{j'} (s,\tau)$, satisfying commutation relations
\begin{align}\label{SU2cr1}
\left[\hat{\cal S}_{0},\hat{\cal S}_{i}\right]=0,\,\hspace{0.5 cm}\,\left[\hat{\cal S}_{i},\hat{\cal S}_{j}\right]=2i\epsilon_{ijk}\frac{\hat{\cal S}_{k}}{n_{0}}.
\end{align}
Based on the results given above, we obtain
\begin{subequations}\label{newsto}
\begin{align}
\hat{\cal S}_{0} &=2A_{0}+\sqrt{2A_{0}}\left(\cos\vartheta\hat{Q}^{(a)}_{1}
+2\sin\vartheta\hat{Q}^{(b)}_{1}\right),\\
\hat{\cal S}_{1}&=2A_{0}\sin(2\vartheta)+\sqrt{2A_{0}}\left(\sin\vartheta\hat{Q}^{(a)}_{1}
+2\cos\vartheta\hat{Q}^{(b)}_{1}\right),\\
\hat{\cal S}_{2}&=\sqrt{2A_{0}}\left(2\cos\vartheta\hat{P}^{(b)}_{1}-2\sin\vartheta
\hat{P}^{(a)}_{1}\right),\\
\hat{\cal S}_{3}&=2A_{0}\cos(2\vartheta)+ \sqrt{2A_{0}}\left(\cos\vartheta\hat{Q}^{(a)}_{1}
-2\sin\vartheta\hat{Q}^{(b)}_{1}\right).
\end{align}
\end{subequations}
We thus have $\langle \hat{\cal S}_0 \rangle=2A_0$.
This is the dimensionless energy of the input probe pulse, independent of the value of $\vartheta$  [the parameter describing the amplitude ration between the two polarization components; see (\ref{CSoliton0})].

If the input probe pulse is circularly polarized, i.e. $\vartheta=0$ (which corresponds to the case where only a single EIT that involves the levels $|1\rangle$ $|4\rangle$, $|3\rangle$ plays a role in the system), one has
\begin{align}\label{Mean1}
\langle\hat{\cal S}_{1}\rangle=\langle\hat{\cal S}_{2}\rangle=0,\,\,\langle\hat{\cal S}_{3}\rangle=\langle\hat{\cal S}_{0}\rangle=2A_{0}.
\end{align}
If the input probe pulse is linearly polarized, i.e. $\vartheta=\pi/4$ (which corresponds to the case where the two EITs involving respectively the levels $|1\rangle$ $|4\rangle$, $|3\rangle$ and the levels $|2\rangle$ $|4\rangle$, $|3\rangle$ play roles simultaneously),
we have
\begin{align}\label{Mean2}
\langle\hat{\cal S}_{2}\rangle=\langle\hat{\cal S}_{3}\rangle=0,\,\,\langle\hat{\cal S}_{1}\rangle=\langle\hat{\cal S}_{0}\rangle=2A_{0}.
\end{align}

Note that to make the two EITs in the system be symmetric and hence the description of the Manakov equations (\ref{QManakoveqs}) be valid, an equal initial ground-state population $S_{11}^{(0)}$ and $S_{22}^{(0)}$ and equal amplitude of the two polarization components of the VOS (\ref{CSoliton0}) must be chosen. This can be realized by taking $S_{11}^{(0)}=S_{22}^{(0)}=0.5$, $\vartheta=\pi/4$, and input probe pulse is linearly polarized (which can be taken as a linear composition of the $\sigma^+$ and $\sigma^-$ polarization components; see the description in Sec.~\ref{Sec2A}).

For describing the polarization squeezing of the probe pulse with the reduced Stokes operators defined by (\ref{newsto}), we introduce the following dark-plane operator
\begin{align}\label{sstokes}
\hat{\cal S}_{\theta}=\cos\theta\hat{\cal S}_{2}+\sin\theta\hat{\cal S}_{3},
\end{align}
where $\theta$ is the detection angle in the dark-plane operator relative to $\hat{\cal S}_{2}$. Such a choice of the dark-plane operator is based on the result (\ref{Mean2}), which indicates that $\langle\hat{\cal S}_{1}\rangle$ is along the Stokes axis and
$(\hat{\cal S}_{2}, \hat{\cal S}_{3})$ are conjugate operator pair in the dark plane. The polarization squeezing occurs when
\begin{eqnarray}
\langle(\Delta\hat{\cal S}_{\theta})^{2}\rangle<|\langle \hat{\cal S}_1\rangle|<\langle(\Delta\hat{\cal S}_{\theta+\pi/2})^{2}\rangle.
\end{eqnarray}
The degree (or amount) of polarization squeezing is described by the squeezing ratio
\begin{align}\label{dps}
R_{p}(\theta)=\frac{\langle(\Delta\hat{\cal S}_{\theta})^{2}\rangle}{|\langle\hat{\cal S}_{0}\rangle|/n_{0}}.
\end{align}

\begin{figure}
\centering
\includegraphics[width=0.9\columnwidth]{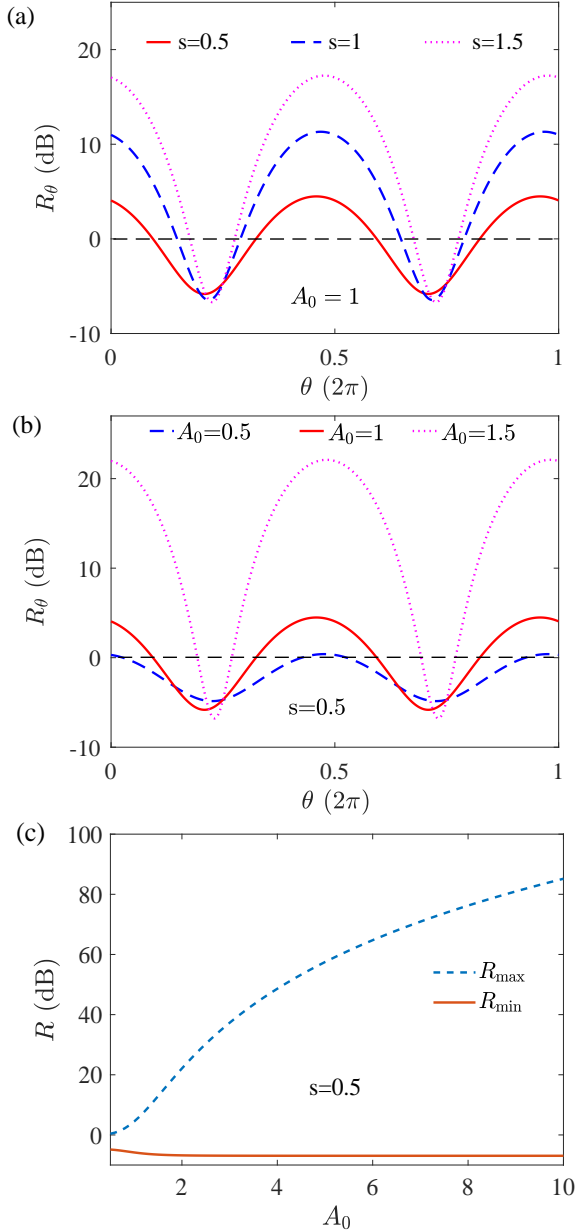}
\caption{
(a)~Polarization squeezing ratio $R_{p}(\theta)$  (in dB) versus detection angle $\theta$ for the dimensionless VOS amplitude $A_{0}=1$ and  different dimensionless propagation distance $s=$0.5 (solid red line), 1 (dashed blue line), and 1.5 (dotted pink line).
(b)~$R_{p}(\theta)$ versus $\theta$ for $s=0.5$ and  different $A_{0}=$0.5 (dashed blue line), 1 (solid red line), and 1.5
(dotted pink line).
(c)~ The maximum ($R_{\rm max}$; dashed blue line) and minimum ($R_{\rm min}$; solid red line) polarization squeezing ratio $R$ versus $A_{0}$ for $s=0.5$.
}
\label{Fig3}
\end{figure}

The variances of $\hat{\cal S}_{\theta}$, i.e. $\langle(\Delta\hat{\cal S}_{\theta})^{2}\rangle$, can be calculated by using (\ref{SolutionofHE}) and (\ref{newsto}) with $\vartheta=\pi/4$. Shown in Fig.~\ref{Fig3}(a)
is the result of numerical simulation on the polarization squeezing ratio $R$ $R_{p}(\theta)$ in decibels (dB), i.e. $10\times\log_{10}R_{p}$,
as a function of the detection angle $\theta$ for the dimensionless VOS amplitude $A_{0}=1$ and different dimensionless propagation distance $s$, with the solid red line, dashed blue line, and dotted pink line being for $s=z/(2L_D)=0.5, 1, 1.5$ (which correspond to $z\approx 1, 2, 3$\,cm), respectively.
We see that the polarization squeezing of the probe pulse indeed occurs in the system, with the squeezing ratio being sensitive to the selection of detection angle $\theta$. Moreover, as the distance $s$ increases, the degree of squeezing  ($\log_{10}R_{p}$ is negative) is increased slowly,
while the degree of anti-squeezing ($\log_{10}R_{p}$ is positive)  grows rapidly.

To explore the property of the polarization squeezing related to the Kerr nonlinearity and the input energy of the probe pulse, the calculation of the squeezing ratio $R_{p}(\theta)$
as a function of $\theta$ is also carried out by fixing $s=0.5$ but varying different VOS amplitude $A_{0}$ (which is proportional to the Kerr nonlinearity and the pulse input energy). Illustrated in Fig.~\ref{Fig3}(b) is the result of the calculation, where the dashed blue line, solid red line, and dashed pink line are for $A_{0}=$0.5, 1, 1.5, respectively. One sees that both the squeezing
and the anti-squeezing grow when  $A_{0}$ is increased (the anti-squeezing grows more faster than the squeezing).
Plotted in Fig.~\ref{Fig3}(c) is the maximum ($R_{\rm max}$;  dashed blue line) and minimum ($R_{\rm min}$; solid red line) polarization squeezing ratio versus $A_{0}$ for $s=0.5$. We see that the squeezing degree of the VOS displays a lower bound (with the value $-6.9$ dB), but the anti-squeezing degree has no upper bound~\cite{note100}.

It should be indicated that, compared with the polarization squeezing of the VOS in optical fibers~\cite{Corney2006,Corney2008}, the polarization squeezing of the VOS in the present atomic system via the DEIT is more efficient. A pronounced feature is that the optical pulse can acquire a large polarization squeezing within a very short propagation distance (in the order of centimeter). The most important physical reason for this is due to the fact that the DEIT-based atomic gas here possesses giant self- and cross-Kerr nonlinearities (which are much larger than that in optical fibers), which make the typical nonlinearity length $L_{NL}$ of the system be small (i.e. order of centimeter). Furthermore, the ultraslow propagating velocity of the VOS is also factor that makes the polarization squeezing more significant.

\begin{figure}
\centering
\includegraphics[width=0.9\columnwidth]{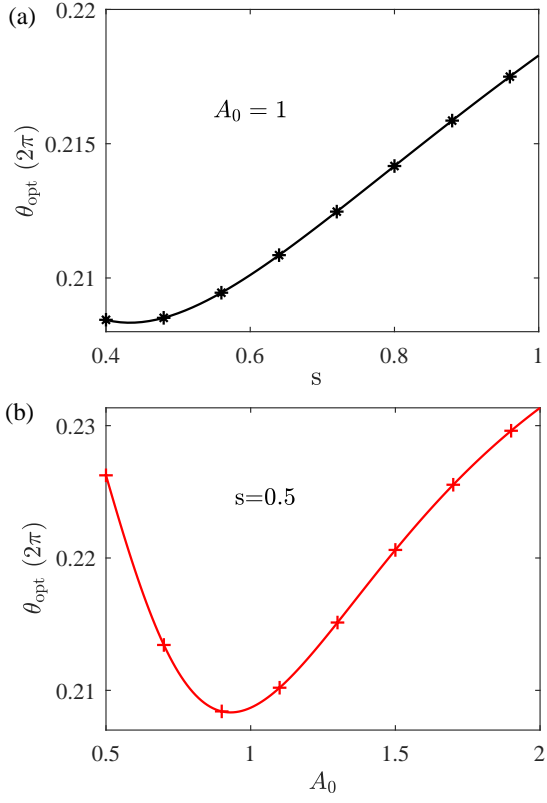}
\caption{
(a)~Numerical result of the optimum detection angle $\theta_{\rm opt}$ for the polarization squeezing as a function of dimensionless propagation distance $s$, for dimensionless amplitude $A_{0}=1$.
(b)~ $\theta_{\rm opt}$ as a function of $A_{0}$ for $s=0.5$.
}
\label{Fig4}
\end{figure}

By minimizing the variance of the dark-plane operator (\ref{sstokes})
with respect to the detection $\theta$, one can obtain the optimum angle $\theta=\theta_{\rm opt}$. Fig.~\ref{Fig4}(a) shows the numerical result of $\theta_{\rm opt}$ as
a function of the propagation distance $s$,
for the dimensionless amplitude $A_{0}=1$; while Fig.~\ref{Fig4}(b) shows $\theta_{\rm opt}$ as a function of $A_{0}$, for $s=0.5$.
Based on these results, experimentally one can choose the optimum detection angle to acquire the largest polarization squeezing of the probe pulse.

\subsection{Atomic spin squeezing}\label{APS}

Due to the significant coupling between the probe pulse and the atoms,
the self- and cross-Kerr nonlinearities in the system not only can result in the large polarization squeezing of the probe pulse (as illustrated above), but also can induce significant spin squeezing of the atoms simultaneously, as shown below.

Because the atoms with the tripod configuration have four levels [see Fig.~\ref{Fig1}(a)], based on the atomic transition operators $\hat{S}_{jl}$ [$j,l=1$-4; see the definition (\ref{Sjl})] one can define fifteen collective spin and multipolar operators [i.e. the generators of SU(4) group] of the atoms, which include three spin operators, five quadrupolar tensor operators, and seven  octupolar tensor operators~\cite{Kikoin2012,Yukawa2016}. These operators are Hermitian ones and hence are observables. One can study the squeezing of these
observables, but here we consider only three classes of spin squeezing for simplicity.

{\bf Class 1}:  This is related to the EIT involving the atomic levels $|1\rangle$, $|4\rangle$, and $|3\rangle$ in Fig.~\ref{Fig1}(a). The three atomic spin operators are given by
\begin{subequations}\label{class1}
\begin{align}
& \hat{J}_{z}=\frac{1}{2}\int d\tau \left[\hat{S}_{11}(\sigma)-\hat{S}_{33}(\sigma)\right],\\
& \hat{J}_{x}=\frac{1}{2}\int d\tau \left[\hat{S}_{13}(\sigma)+\hat{S}_{31}(\sigma)\right],\\
& \hat{J}_{y}=\frac{i}{2}\int d\tau \left[\hat{S}_{13}(\sigma)-\hat{S}_{31}(\sigma)\right].
\end{align}
\end{subequations}

{\bf Class  2}:  This is related to the EIT involving atomic levels $|2\rangle$, $|4\rangle$, and $|3\rangle$. The three atomic spin operators read
\begin{subequations}\label{class2}
\begin{align}
& \hat{J}_{z}=\frac{1}{2}\int d\tau \left[\hat{S}_{22}(\sigma)-\hat{S}_{33}(\sigma)\right],\\
& \hat{J}_{x}=\frac{1}{2}\int d\tau \left[\hat{S}_{23}(\sigma)+\hat{S}_{32}(\sigma)\right],\\
& \hat{J}_{y}=\frac{i}{2}\int d\tau \left[\hat{S}_{23}(\sigma)-\hat{S}_{32}(\sigma)\right].
\end{align}
\end{subequations}

{\bf Class  3}:
If a coherence is initially prepared between the two ground states $|1\rangle$ and $|2\rangle$, i.e. $S_{21}^{(0)}=S_{12}^{(0)}\neq 0$, the system can support another class of atomic spin squeezing, with the spin operators defined by
\begin{subequations}\label{class3}
\begin{align}
& \hat{J}_{z}=\frac{1}{2}\int d\tau \left[\hat{S}_{12}(\sigma)+\hat{S}_{21}(\sigma)\right],\\
& \hat{J}_{x}=\frac{1}{2}\int d\tau \left[\hat{S}_{12}(\sigma)-\hat{S}_{21}(\sigma)\right],\\
& \hat{J}_{y}=\frac{i}{2}\int d\tau \left[\hat{S}_{11}(\sigma)-\hat{S}_{22}(\sigma)\right].
\end{align}
\end{subequations}
It is easy to show that the atomic spin operators defined in these three classes satisfy the commutation relations
\begin{align}\label{spincr}
\left[\hat{J}_{i},\hat{J}_{j}\right]=i\epsilon_{ijk}\hat{J}_{k},
\end{align}
where $i,j,k=x,y,z$.

Because the mean polarization for all three classes is along the direction of $\hat{J}_{z}$ (i.e. the ``Stokes axis'' is the $z$ axis), to estimate the spin squeezing we can introduce the following spin dark-plane operator
\begin{align}\label{spin2}
\hat{J}_{\theta}&=\frac{1}{2}\int d\tau\left[\hat{S}_{\alpha}e^{-i\theta}+\hat{S}_{\beta}
e^{i\theta}\right]\notag\\
&=\cos\theta\,\hat{J}_{x}+\sin\theta\,\hat{J}_{y},
\end{align}
where $(\alpha,\beta)=(31,13)$ is for the class 1,  $(\alpha,\beta)=(32,23)$ is for the class 2 and $(\alpha,\beta)=(21,12)$ is for the class 3.
The degree (amount) of the spin squeezing can be described by
\begin{align}\label{spde}
\xi^{2}=\frac{\langle(\Delta\hat{J}_{\theta})^{2}\rangle}{\langle\hat{J}_{z}\rangle/2}.
\end{align}
If $\xi^{2}<1$, the atomic state is said to be spin squeezed~\cite{Ma2011}.
From the results given in the Appendix~\ref{app2} (where the dynamical equations for $\hat{S}_{jl}$ and $\hat{E}_{pj}$ are solved simultaneously for all atomic levels; no adiabatical elimination of the upper levels is used), especially (\ref{S31}), we can calculate the amount of spin squeezing
based on the solution on $\hat{E}_{pj}$  given above.

\begin{figure}
\centering
\includegraphics[width=0.95\columnwidth]{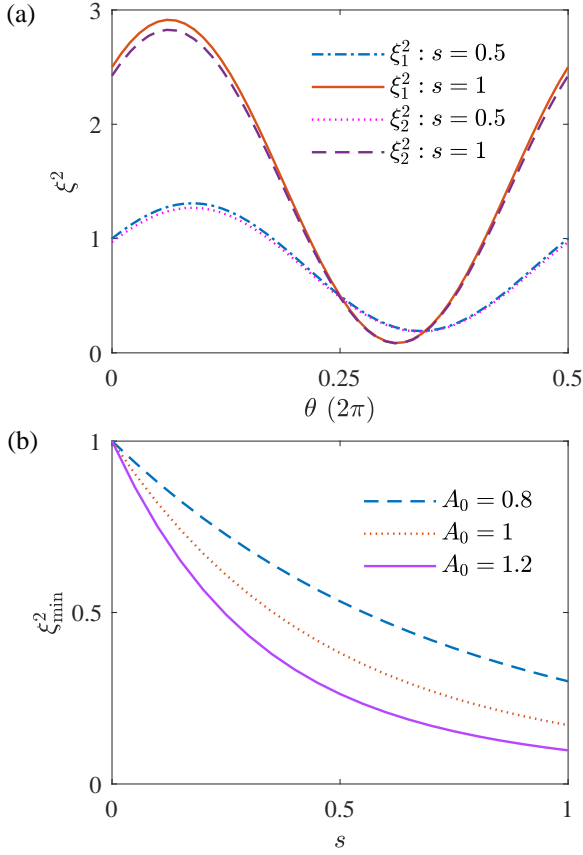}
\caption{
(a)~The amount of atomic spin squeezing  $\xi_{1}^{2}$ and $\xi_{2}^{2}$ (for the class 1 and class 2, respectively) as functions of detection angle $\theta$, for dimensionless VOS amplitude $A_{0}=1$ and  dimensionless propagation distances $s=0.5$ (dot-dashed blue line for the class 1, dotted pink line for the class 2), and $s=1$ (solid orange line for the class 1, dashed purple line for the class 2).
(b)~Minimum atomic spin squeezing degree $\xi^{2}_{\rm min}$ of the class 1 as a function of $s$, for $A_{0}=0.8$ (dashed blue line), 1 (dotted orange line), and 1.2 (solid purple line).
}
\label{Fig5}
\end{figure}

Shown in Fig.~\ref{Fig5}(a)
are results on the amount of atomic spin squeezing  $\xi_{1}^{2}$ (for the class 1) and $\xi_{2}^{2}$ (for the class 2) as functions of detection angle $\theta$, by taking the dimensionless VOS amplitude $A_{0}=1$. In the figure, the dot-dashed blue line (dotted pink line) is for the class 1 (class 2) for $s=0.5$, while the solid orange line (dashed purple line) is for the class 1  (class 2) for $s=1$.
We see that the system indeed supports significant squeezing of the atomic spins, which can reach a minimum value by choosing the value of $\theta$.  The spin squeezing of both class 1 and class 2 are nearly the same, this is due to the fact that the two EITs configurations in the system are highly symmetric.

To reveal the relation between the spin squeezing and the Kerr nonlinearity of the system, a calculation on the minimum spin squeezing degree $\xi^{2}_{\rm min}$ of the class 1  as a function of $s$ is carried out, with the result plotted in Fig.~\ref{Fig5}(b). In the figure, the dashed blue line, dotted orange line, and solid purple line are for $A_{0}=0.8$, 1, and 1.2, respectively. One sees that as $A_{0}$ and $s$ increase, $\xi^{2}_{\rm min}$ is reduced. This means that the stronger the Kerr nonlinearity, the larger the spin squeezing. This conclusion can also be obtained through the calculation of the minimum spin squeezing degree $\xi^{2}_{\rm min}$ of the class 2.

One can also obtain the degree of atomic spin squeezing  $\xi_{3}^{2}$ for the class 3  as a function of $\theta$ and $s$ with different $A_{0}$. The result shows that the behavior of $\xi_3^{2}$ is similar to that of the polarization squeezing degree $R_{\theta}$ of the probe pulse (i.e. Fig.~\ref{Fig3}); see Appendix~\ref{app4} for details.

From the above results we see that the spin squeezing of the atoms occurs simultaneously with the polarization squeezing shown in the last subsection, both of which originate from the giant self- and cross-Kerr nonlinearities resulted from the perturbed DEIT; during the formation the simultaneous squeezing of light polarization and the atomic spins, the zero modes of the quantum fluctuations  in the system play very important roles.

\section{Discussion and summary}\label{Sec5}

Since the atomic gas we consider is dilute, the direct interaction between atoms can be neglected. If the system works under the condition of strict DEIT (i.e. $\Delta_3=0$), no squeezing occurs both for the probe pulse and for the atoms. The physical reasons for the occurrence of the simultaneous squeezing of the probe pulse and the atomic spins described above can be understood as follows. (i) The use of a perturbed DEIT brings a significant coupling between the probe pulse and the atoms, which makes the system have giant Kerr nonlinear and second-order dispersion effects that can result in the formation of the VOS and the polarization squeezing of the probe pulse. (ii) Under the condition of the perturbed DEIT, the significant coupling between the probe pulse and the atoms induces an indirect interaction between the atoms, and hence the atomic spin squeezing can be generated simultaneously with the appearance of the squeezing of the probe pulse.

In conclusion, we have investigated the quantum dynamics of a weakly nonlinear probe pulse with two polarization components, which is coupled to a cold atomic ensemble and working under the condition of perturbed DEIT. We have derived two coupled quantum NLS equations from the MHL equations and developed a quantum theory of VOS, which have ultraslow propagation velocity and extremely low generation power. We have solved the non-Hermitian eigenvalue problem that describes the quantum fluctuations on the VOS background, and rigorously proved that all fluctuation eigenmodes (including the continuous modes and the zero modes) obtained constitute a bi-orthonormal and complete set. We have found that, due to the giant self- and cross-Kerr nonlinearities contributed by the DEIT, a significant polarization squeezing of the probe pulse can be realized in the system. We have also found that a large squeezing of atomic spins can be generated, which appears simultaneously with the occurrence of the polarization squeezing of the probe pulse. The zero modes of the quantum fluctuations are the main origin  for the formation of such a simultaneous squeezing.

The remarkable conclusions for generating the simultaneous squeezing of light polarizations and atomic spins by using only a coherent probe pulse obtained here opens a way for revealing the unique property of quantum squeezing in coupled light-atom systems, and also for promising applications in quantum information and precision measurement.

\section*{Acknowledgements}
This work was supported by the National Natural Science Foundation of China under Grant No.~11975098, and by the Research Funds of Happiness Flower ECNU under Grant No.~2020ECNU-XFZH005.

\appendix

\section{Explicit expressions of the Heisenberg-Langevin equations}\label{app1}

Explicit forms of the Heisenberg-Langevin equations~(\ref{HLM0}) read
\begin{widetext}
\begin{subequations}\label{HLEqs}
\begin{align}
& i\left(\frac{\partial}{\partial t}+\Gamma_{31}\right)\hat{S}_{11}-i\Gamma_{13}\hat{S}_{33}-i\Gamma_{14}\hat{S}_{44}+g_{p1}^\ast{\hat E}_{p1}^{\dag}\hat{S}_{41}-g_{p1}\hat{S}_{14}{\hat E}_{p1}-i{\hat F}_{11}=0,\\
& i\left(\frac{\partial}{\partial t}+\Gamma_{32}\right)\hat{S}_{22}-i\Gamma_{23}\hat{S}_{33}-i\Gamma_{24}\hat{S}_{44}+g_{p2}^\ast{\hat E}_{p2}^{\dag}\hat{S}_{42}-g_{p2}\hat{S}_{24}{\hat E}_{p2}-i{\hat F}_{22}=0,\\
& i\left(\frac{\partial}{\partial t}+\Gamma_{3}\right)\hat{S}_{33}-i\Gamma_{31}\hat{S}_{11}-i\Gamma_{32}\hat{S}_{22}-i\Gamma_{34}\hat{S}_{44}+\Omega_{c}^\ast\hat{S}_{43}-\Omega_{c}\hat{S}_{34}-i{\hat F}_{33}=0,\\
& i\left(\frac{\partial}{\partial t}+\Gamma_{4}\right)\hat{S}_{44}-\Omega_{c}^\ast\hat{S}_{43}+\Omega_{c}\hat{S}_{34}-g_{p1}^\ast{\hat E}_{p1}^{\dag}\hat{S}_{41}+g_{p1}\hat{S}_{14}{\hat E}_{p1}-g_{p2}^\ast{\hat E}_{p2}^{\dag}\hat{S}_{42}+g_{p2}\hat{S}_{24}{\hat E}_{p2}-i{\hat F}_{44}=0,\\
& \left(i\frac{\partial}{\partial t}+d_{21}\right)\hat{S}_{21}+g_{p2}^{\ast}{\hat E}_{p2}^{\dag}\hat{S}_{41}-g_{p1}\hat{S}_{24}{\hat E}_{p1}-i{\hat F}_{21}=0, \\
& \left(i\frac{\partial}{\partial t}+d_{43}\right)\hat{S}_{43}+\Omega_{c}\left(\hat{S}_{33}-\hat{S}_{44}\right)+g_{p1+}\hat{S}_{13}{\hat E}_{p1}+g_{p2}\hat{S}_{23}{\hat E}_{p2}-i{\hat F}_{43}=0,\\
& \left(i\frac{\partial}{\partial t}+d_{31}\right)\hat{S}_{31}+\Omega_{c}^\ast\hat{S}_{41}-g_{p1}\hat{S}_{34}{\hat E}_{p1}-i{\hat F}_{31}=0, \\
& \left(i\frac{\partial}{\partial t}+d_{32}\right)\hat{S}_{32}+\Omega_{c}^\ast\hat{S}_{42}-g_{p2}\hat{S}_{34}{\hat E}_{p2}-i{\hat F}_{32}=0, \\
& \left(i\frac{\partial}{\partial t}+d_{41}\right)\hat{S}_{41}+\Omega_{c}\hat{S}_{31}+g_{p1}\left(\hat{S}_{11}-\hat{S}_{44}\right){\hat E}_{p1}+g_{p2}\hat{S}_{21}{\hat E}_{p2}-i{\hat F}_{41}=0, \\
& \left(i\frac{\partial}{\partial t}+d_{42}\right)\hat{S}_{42}+\Omega_{c}\hat{S}_{32}+g_{p2}\left(\hat{S}_{22}-\hat{S}_{44}\right){\hat E}_{p2}+g_{p1}\hat{S}_{12}{\hat E}_{p1}-i{\hat F}_{41}=0.
\end{align}
\end{subequations}
\end{widetext}
Here $d_{\alpha\beta}=\Delta_{\alpha}-\Delta_{\beta}+i\gamma_{\alpha\beta}$
($\alpha\neq \beta)$ with $\gamma_{\alpha\beta}\equiv(\Gamma_\alpha+\Gamma_\beta)/2+\gamma_{\alpha\beta}^{\rm dep}$, $\Gamma_\beta\equiv\sum_{\alpha<\beta}\Gamma_{\alpha\beta}$, and $\Gamma_{\alpha\beta}$ is the decay rate of the spontaneous emission from the state $|\beta\rangle$ to the state $|\alpha\rangle$, $\gamma_{\alpha\beta}^{\rm dep}$ is the dephasing rate between $|\alpha\rangle$ and $|\beta\rangle$.
The two-time correlation functions of ${\hat F}_{\alpha\beta}$ are given by
$\langle{\hat F}_{\alpha\beta}(z,t){\hat F}_{\alpha'\beta'}(z',t')\rangle\equiv{\rm Tr}_{\rm R}[{\hat F}_{\alpha\beta}(z,t){\hat F}_{\alpha'\beta'}(z',t'){\hat S }_{\rm R}]$, where ${\hat S }_{\rm R}$ is the initial density operator of the thermal reservoir coupling to the atomic system, ${\rm Tr}_{\rm R}$ denotes the trace over the reservoir variables.

\section{Derivation of the coupled quantum NLS equations}\label{app2}

Due to the difficulties for solving quantum nonlinear problems, up to now there is no quantum reductive perturbation method developed by which one can derive a quantum NLS equation directly from coupled nonlinear quantum partial differential equations involving many degrees of freedom of both atoms and quantized light fields.
Here, we give a heuristic derivation on the coupled quantum NLS Eqs.~(\ref{CQNLS}) describing the nonlinear evolution of the probe-field envelope $\hat{E}_{pj}$ in the present system. The derivation can be divided into two steps.

{\it Step 1:  Quantum linear Schr\"odinger equation with group-velocity dispersion.}  We assume that the probe field is very weak so that the Kerr nonlinearity in the system can be neglected.
Thus the Heisenberg-Langevin and Maxwell equations can be treated by using a linear approximation.
By taking ${\hat S}_{\alpha\beta}\rightarrow S_{\alpha\beta}^{(0)}+{\hat S}_{\alpha\beta}$. Here $S_{\alpha\beta}^{(0)}$ is the steady-state solution of ${\hat S}_{\alpha\beta}$ when the probe pulse is not applied (when i.e.  $\hat{E}_{pj}=0$), satisfying
\begin{eqnarray}
&& S_{11}^{(0)}+S_{22}^{(0)}=1,
\end{eqnarray}
with $S_{12}^{(0)}=S_{21}^{(0)}$ arbitrary and all other $S_{\alpha\beta}^{(0)}=0$. In order to have a symmetry between the two EITs in Fig.~\ref{Fig1}(a), we take $S_{11}^{(0)}=S_{22}^{(0)}=0.5$.
We then obtain
the linearized equations of Eqs.~(\ref{HLM}), which can be solved by using a Fourier transform. After eliminating the atomic variables, we obtain
\begin{align}\label{LSE}
\left[i\frac{\partial}{\partial z}+K_{j}(\omega)\right]{\tilde{\hat E}}_{pj}(z,\omega)=i{\tilde{\hat{{\cal F}}}}_{pj}(z,\omega),
\end{align}
$j=1, 2$. Here $\omega$ is the sideband frequency of the probe pulse, ${\tilde{\hat E}}_{pj}(z,\omega)$ and $\tilde{\hat{{\cal F}}}_{pj}(z,\omega)$
are respectively the Fourier transforms of $\hat{E}_{pj}(z,t)$ and  $\hat{\cal F}_{pj}(z,t)$, and $K_{j}$ is the linear dispersion relation defined by
\begin{align}
&K_{j}(\omega)=\frac{\omega}{c}+\frac{|g_{pj}|^{2}N}{c}\frac{(\omega+d_{3j}) S_{jj}^{(0)}}{D_{j}(\omega)},
\end{align}
The new noise operator $\hat{{\cal F}}_{p}(z,t)$ is defined by
\begin{align}
\hat{{\cal F}}_{pj}(z,t)=\frac{ g_{pj}^\ast N}{c}\frac{\left(\omega+d_{3j}\right) \hat{F}_{4j}(z,t)-\Omega_{c}\hat{F}_{3j}(z,t)}{D(\omega)},
\end{align}
where $D_{j}(\omega)=|\Omega_{c}|^2-(\omega+d_{3j})(\omega+d_{4j})$.

Assuming that the bandwidth of the probe pulse is not too narrow, one can expand $K_{j}(\omega)$ in a Taylor series around $\omega=0$ up to the second-order in $\omega$, i.e. $K(\omega)_{j}\approx K_{0j}+\omega/V_{gj}+K_{2j}\omega^2/2$. Here $K_{0j}\equiv K_{j}|_{\omega=0}$, $V_{gj}^{-1}\equiv K_{1j}\equiv(\partial K_{j}/\partial\omega)|_{\omega=0}$ is the group-velocity dispersion of the probe field, and $K_{2j}\equiv(\partial^2K_{j}/\partial\omega^2)|_{\omega=0}$ is the coefficient denoting the group-velocity dispersion. Substituting this expansion into the envelope equation (\ref{LSE}) and convert it back to time domain by using an inverse Fourier transformation, we arrive the quantum linear Schr\"odinger equation
\begin{equation}\label{Linear Eq}
i\left(\frac{\partial}{\partial z}+\frac{1}{V_{gj}}\frac{\partial}{\partial t}\right)\hat{E}_{pj}+K_{0j}\hat{E}_{pj}-\frac{K_{2j}}{2}\frac{\partial^2}{\partial t^2}\hat{E}_{pj}=i{\hat{\cal F}}_{pj},
\end{equation}
where ${\hat{\cal F}}_{pj}(z,t)$ is the inverse Fourier transform of ${\tilde{\hat{\cal F}}}_{pj}(z,\omega)$.

{\it Step 2: Quantum nonlinear equation with cubic Kerr nonlinearity.} We next derive the equation for a weakly-nonlinear probe field for which the group-velocity dispersion can be neglected but the Kerr-nonlinearity is considered. This is valid when the probe pulse has a long-time duration, so that the time derivatives in the HLM Eqs.~(\ref{HLM}) play negligible roles.
To get the equation for $\hat{E}_{pj}$ we employ an iteration method by taking $g_{pj}\hat{E}_{pj}$ as a small quantity.
By considering the steady-state solution  of the Heisenberg-Langevin equations, we obtain the solution at the zero-order approximation, given by
\begin{subequations}\label{it0}
\begin{align}
S_{44}^{(0)}&=\frac{\Gamma_{31}\Gamma_{32}B_{1}}{A_{1}B_{2}+B_{1}A_{2}},\\
S_{33}^{(0)}&=\frac{A_{1}S_{44}^{(0)}}{B_{1}},\\
S_{22}^{(0)}&=\frac{\Gamma_{23}S_{33}^{(0)}+\Gamma_{24}S_{44}^{(0)}}{\Gamma_{32}},\\
S_{11}^{(0)}&=\frac{\Gamma_{13}S_{33}^{(0)}+\Gamma_{14}S_{44}^{(0)}}{\Gamma_{31}},\\
S_{43}^{(0)}&=\frac{\Omega_{c}\left(S_{44}^{(0)}-S_{33}^{(0)}\right)}{d_{43}},
\end{align}
\end{subequations}
with $S_{12}^{(0)}=S_{21}^{(0)}$ arbitrary and other $S_{\alpha\beta}^{(0)}=0$. Here
\begin{subequations}
\begin{align}
A_{1}&=i\Gamma_{4}+|\Omega_{c}|^{2}(1/d_{43}^{\ast}-1/d_{43}),\\
B_{1}&=|\Omega_{c}|^{2}(1/d_{43}^{\ast}-1/d_{43}),\\
A_{2}&=\Gamma_{31}\Gamma_{24}+\Gamma_{32}\Gamma_{14}+\Gamma_{32}\Gamma_{31},\\
B_{2}&=\Gamma_{31}\Gamma_{32}+\Gamma_{32}\Gamma_{13}+\Gamma_{23}\Gamma_{31}.
\end{align}
\end{subequations}

The first-order solution reads
\begin{align}\label{S31}
\hat{S}_{\alpha j}^{(1)}=a_{\alpha j}^{(1)}g_{pj}\hat{E}_{pj}\,(\alpha=3,4, j=1,2),
\end{align}
with other $\hat{S}_{\alpha\beta}^{(1)}=0$, and
\begin{subequations}\label{it1}
\begin{align}
a_{3j}^{(1)}&=\frac{\Omega_{c}^{\ast}\left(S_{44}^{(0)}
-S_{jj}^{(0)}\right)-d_{4j}S_{43}^{\ast(0)}}{X_{j}},\\
a_{4j}^{(1)}&=\frac{d_{3j}\left(S_{jj}^{(0)}
-S_{44}^{(0)}\right)+\Omega_{c}S_{43}^{\ast(0)}}{X_{j}},
\end{align}
\end{subequations}
with $X_{j}=|\Omega_{c}|^2-d_{3j}d_{4j}$.

The second-order solution reads
\begin{subequations}\label{it2}
\begin{align}
\hat{S}_{21}^{(2)}&=a_{21}^{(2)}g_{p1}g_{p2}^{\ast}\hat{E}_{p2}^{\dagger}\hat{E}_{p1},\\
\hat{S}_{44}^{(2)}&=a_{441}^{(2)}|g_{p1}|^{2}\hat{E}_{p1}^{\dagger}\hat{E}_{p1}+a_{442}^{(2)}|g_{p2}|^{2}\hat{E}_{p2}^{\dagger}\hat{E}_{p2},\\
\hat{S}_{33}^{(2)}&=a_{331}^{(2)}|g_{p1}|^{2}\hat{E}_{p1}^{\dagger}\hat{E}_{p1}+a_{332}^{(2)}|g_{p2}|^{2}\hat{E}_{p2}^{\dagger}\hat{E}_{p2},\\
\hat{S}_{22}^{(2)}&=a_{221}^{(2)}|g_{p1}|^{2}\hat{E}_{p1}^{\dagger}\hat{E}_{p1}+a_{222}^{(2)}|g_{p2}|^{2}\hat{E}_{p2}^{\dagger}\hat{E}_{p2},\\
\hat{S}_{11}^{(2)}&=a_{111}^{(2)}|g_{p1}|^{2}\hat{E}_{p1}^{\dagger}\hat{E}_{p1}+a_{112}^{(2)}|g_{p2}|^{2}\hat{E}_{p2}^{\dagger}\hat{E}_{p2},\\
\hat{S}_{43}^{(2)}&=a_{431}^{(2)}|g_{p1}|^{2}\hat{E}_{p1}^{\dagger}\hat{E}_{p1}+a_{432}^{(2)}|g_{p2}|^{2}\hat{E}_{p2}^{\dagger}\hat{E}_{p2},
\end{align}
\end{subequations}
with other $\hat{S}_{\alpha\beta}^{(2)}=0$, and
\begin{subequations}
\begin{align}
a_{21}^{(2)}&=\frac{a_{42}^{\ast(1)}-a_{41}^{(1)}}{d_{21}},\\
a_{441}^{(2)}&=\frac{B_{2}s_{11}+B_{1}s_{21}}{A_{1}B_{2}+B_{1}A_{2}},\\
a_{442}^{(2)}&=\frac{B_{2}s_{12}+B_{1}s_{22}}{A_{1}B_{2}+B_{1}A_{2}},\\
a_{331}^{(2)}&=\frac{A_{1}s_{21}-A_{2}s_{11}}{A_{1}B_{2}+B_{1}A_{2}},\\
a_{332}^{(2)}&=\frac{A_{1}s_{22}-A_{2}s_{12}}{A_{1}B_{2}+B_{1}A_{2}},\\
a_{221}^{(2)}&=\frac{\Gamma_{23}a_{331}^{(2)}+\Gamma_{24}a_{441}^{(2)}}{\Gamma_{32}},\\
a_{222}^{(2)}&=\frac{\Gamma_{23}a_{332}^{(2)}+\Gamma_{24}a_{442}^{(2)}+i(a_{42}^{(1)}-a_{42}^{\ast(1)})}{\Gamma_{32}},\\
a_{111}^{(2)}&=\frac{\Gamma_{13}a_{331}^{(2)}+\Gamma_{14}a_{441}^{(2)}+i(a_{41}^{(1)}-a_{41}^{\ast(1)})}{\Gamma_{31}},\\
a_{112}^{(2)}&=\frac{\Gamma_{13}a_{332}^{(2)}+\Gamma_{14}a_{442}^{(2)}}{\Gamma_{32}},\\
a_{431}^{(2)}&=\frac{\Omega_{c}(a_{441}^{(2)}-a_{331}^{(2)})-a_{31}^{\ast(1)}}{d_{43}},\\
a_{432}^{(2)}&=\frac{\Omega_{c}(a_{442}^{(2)}-a_{332}^{(2)})-a_{32}^{\ast(1)}}{d_{43}},
\end{align}
\end{subequations}
here
\begin{subequations}
\begin{align}
s_{11}&=\Omega_{c}a_{31}^{(1)}/d_{43}^{\ast}-\Omega_{c}^{\ast}a_{31}^{\ast(1)}/d_{43}+a_{41}^{(1)}-a_{41}^{\ast(1)},\\
s_{12}&=\Omega_{c}a_{32}^{(1)}/d_{43}^{\ast}-\Omega_{c}^{\ast}a_{32}^{\ast(1)}/d_{43}+a_{42}^{(1)}-a_{42}^{\ast(1)},\\
s_{21}&=i\Gamma_{32}(a_{41}^{\ast(1)}-a_{41}^{(1)}),\\
s_{22}&=i\Gamma_{31}(a_{42}^{\ast(1)}-a_{42}^{(1)}).
\end{align}
\end{subequations}

Proceeding to the third order, we obtain
\begin{subequations}\label{it3}
\begin{align}
\hat{S}_{41}^{(3)}&=a_{411}^{(3)}|g_{p1}|^{2}g_{p1}\hat{E}_{p1}^{\dagger}\hat{E}_{p1}\hat{E}_{p1}\notag\\
&\hspace{0.5 cm}+a_ {412}^{(3)}|g_{p2}|^{2}g_{p1}\hat{E}_{p2}^{\dagger}\hat{E}_{p2}\hat{E}_{p1},\\
\hat{S}_{42}^{(3)}&=a_{421}^{(3)}|g_{p1}|^{2}g_{p2}\hat{E}_{p1}^{\dagger}\hat{E}_{p1}\hat{E}_{p2}\notag\\
&\hspace{0.5 cm}+a_{422}^{(3)}|g_{p2}|^{2}g_{p2}\hat{E}_{p2}^{\dagger}\hat{E}_{p2}\hat{E}_{p2},
\end{align}
\end{subequations}
and
\begin{subequations}
\begin{align}
a_{411}^{(3)}=&\frac{\Omega_{c}a_{431}^{\ast(2)}-d_{31}(a_{441}^{(2)}-a_{111}^{(2)})}{X_{1}},\\
a_{412}^{(3)}=&\frac{\Omega_{c}a_{432}^{\ast(2)}-d_{31}(a_{442}^{(2)}-a_{112}^{(2)}+a_{21}^{(2)})}{X_{1}},\\
a_{422}^{(3)}=&\frac{\Omega_{c}a_{432}^{\ast(2)}-d_{32}(a_{442}^{(2)}-a_{222}^{(2)})}{X_{2}},\\
a_{421}^{(3)}=&\frac{\Omega_{c}a_{431}^{\ast(2)}-d_{32}(a_{441}^{(2)}-a_{221}^{(2)}+a_{21}^{\ast(2)})}{X_{2}}.
\end{align}
\end{subequations}
The solutions of other ${\hat S}_{\alpha\beta}$ are also obtained but are omitted here. Exact to the third-order approximation with respect to $g_{p}\hat{E}_{p}$, we obtain the perturbation expansion of ${\hat S}_{4j}$, given by
\begin{equation}\label{S4j}
{\hat S}_{4j}={\hat S}_{4j}^{(1)}+{\hat S}_{4j}^{(3)},
\end{equation}
Here the first (second) term on the right hand side of the above expression describes the linear (nonlinear) response of the atoms to the probe field.
Substituting Eq.~(\ref{S4j}) into Eq.~(\ref{HLM}b), we arrive at the nonlinear equation
\begin{align}\label{Nonlinear Eq}
&\left(i\frac{\partial}{\partial z}+K_{0j}\right)\hat{E}_{pj}+\notag\\
&(W_{jj}|g_{pj}|^2\hat{E}_{pj}^{\dag}\hat{E}_{pj}+W_{j3-j}|g_{p3-j}|^2\hat{E}_{p3-j}^{\dag}\hat{E}_{p3-j})\hat{E}_{pj}=0,
\end{align}
where the coefficients  of the self-phase and cross-phase modulations  appearing in the above equation, given by
\begin{subequations}\label{nlcoeff}
\begin{align}
W_{11}&=\frac{N|g_{p1}|^{2}}{c}a_{411}^{(3)},\\
W_{12}&=\frac{N|g_{p2}|^{2}}{c}a_{412}^{(3)},\\
W_{22}&=\frac{N|g_{p2}|^{2}}{c}a_{422}^{(3)},\\
W_{21}&=\frac{N|g_{p1}|^{2}}{c}a_{421}^{(3)},
\end{align}
\end{subequations}
and satisfying the relation $W_{11}W_{22}=W_{12}W_{21}$.
By combining Eqs.~(\ref{Linear Eq}) and~(\ref{Nonlinear Eq}), we obtain the coupled quantum NLS equations  for $\hat{E}_{pj}$:
\begin{align}\label{QNLSE}
&\left[i\left(\frac{\partial}{\partial z}+\frac{1}{V_{gj}}\frac{\partial}{\partial t}\right)\right]\hat{E}_{pj}-\frac{K_{2j}}{2}\frac{\partial^2}{\partial t^2}\hat{E}_{pj}+(W_{jj}|g_{pj}|^2\hat{E}_{pj}^{\dag}\hat{E}_{pj}\notag\\
&+W_{j3-j}|g_{p3-j}|^2\hat{E}_{p3-j}^{\dag}\hat{E}_{p3-j})\hat{E}_{pj}=i{\hat{\cal F}}_{pj},
\end{align}
which is valid for probe fields when the group-velocity dispersion and cubic Kerr nonlinearity play equal roles. By making the transformation $\hat{E}_{pj}\rightarrow\hat{E}_{pj}\exp[i{\rm Re}(K_{0j})z]$, the above equation becomes the coupled quantum NLS Eqs.~(\ref{CQNLS}) given in the main text.

Notice that, under the DEIT condition, the Langevin noise ${\hat{\cal F}}_{pj}$ plays a negligible role in the system.  The reason is that, at an ultracold environment, the excitation energy of probe photons, i.e. $\hbar\omega_{p}$, is much larger than that of the thermal noises, which is of order $k_{\rm B}T$ (here $k_{\rm B}$ is the Boltzmann constant and $T$ is temperature). Thus the average number of thermal noise photons, i.e.   i.e. ${\bar n}_{\rm th}\equiv\{\exp[\hbar\omega_{p}/(k_{B}T)]-1\}^{-1}$, is vanishing small. Thus
the thermal reservoir coupling to the atomic medium can be safely regarded as a vacuum reservoir ${\hat\rho}_{\rm R}\approx|\{0\}_{\rm R}\rangle\langle\{0\}_{\rm R}|$~\cite{Zhang2021,Gorshkov2007PRA1,Gorshkov2007PRA2}.  In addition, the atomic population at the excited state $|4\rangle$ is always very small due to the EIT effect, which make the spontaneous emission of the atoms  (and hence the dissipation of the probe pulse during propagation) be suppressed greatly.
As a result, the Langevin noise operators make negligible contributions to all normally-ordered two-time correlation functions~\cite{Zhang2021,Gorshkov2007PRA1,Gorshkov2007PRA2}.


\section{Simplification of the envelope Eq.~(\ref{dml})}\label{app3}

The dimensionless form of the coupled quantum NLS Eqs.~(\ref{dml}) read as
\begin{align}\label{c1}
&i\left(\frac{\partial}{\partial s}+2\alpha_{j}\right)\hat{U}_{j}+ig_{\delta}\frac{\partial}{\partial\tau}\hat{U}_{1}+g_{Dj}\frac{\partial^2}{\partial \tau^2}\hat{U}_{j}\notag\\
&\hspace{1 cm}+2\left(g_{jj}\hat{U}_{j}^{\dag}\hat{U}_{j}+g_{j3-j}
\hat{U}_{3-j}^{\dagger}\hat{U}_{3-j}\right)\hat{U}_{j}=0.
\end{align}
Let $s=s_{0}s'$, $\tau=\tau_{0}\tau'$, $\hat{U}_{j}=U_{0}\hat{U}_{j}'$,  then Eq.~(\ref{c1}) can be written as
\begin{align}\label{c2}
&i\left(\frac{1}{s_{0}}\frac{\partial}{\partial s'}+2\alpha_{j}\right)U_{0}\hat{U}'_{j}+i\frac{g_{\delta}U_{0}}{\tau_{0}}\frac{\partial}{\partial\tau'} \hat{U}'_{j}+\frac{g_{Dj}U_{0}}{\tau_{0}^{2}}\frac{\partial^2}{\partial\tau'^2}\hat{U}_{j}\notag\\
&+2\frac{U_{0}^{3}}{g_{11}}\left(\frac{g_{jj}}{g_{11}}\hat{U}_{j}^{\prime\dag}\hat{U}'_{j}+\frac{g_{j3-j}}{g_{11}}
\hat{U}_{3-j}^{\prime\dagger}\hat{U}'_{3-j}\right)\hat{U}'_{j}=0,
\end{align}
divide both sides by $U_{0}/s_{0}$, Eq.~(\ref{c2}) turns into
\begin{align}\label{c3}
&i\left(\frac{\partial}{\partial s'}+2\alpha_{j}s_{0}\right)\hat{U}'_{j}+i\frac{g_{\delta}s_{0}}{\tau_{0}}\frac{\partial}{\partial\tau'} \hat{U}'_{j}+\frac{g_{Dj}s_{0}}{\tau_{0}^{2}}\frac{\partial^2}{\partial\tau'^2}\hat{U}'_{j}\notag\\
&+2\frac{U_{0}^{2}s_{0}}{g_{11}}\left(\frac{g_{jj}}{g_{11}}\hat{U}_{j}^{\prime\dag}\hat{U}'_{j}+\frac{g_{j3-j}}{g_{11}}
\hat{U}_{3-j}^{\prime\dagger}\hat{U}'_{3-j}\right)\hat{U}'_{j}=0.
\end{align}
Due to the symmetry of the system, we have $g_{D1}\approx g_{D2}=g_{D}$, $g_{11}\approx g_{22}$, by setting $g_{D}s_{0}/\tau_{0}^{2}=1$ and $U_{0}^{2}s_{0}/g_{11}=1$, then Eq.~(\ref{c3}) can be reduced into the perturbed quantum Manakov equations
\begin{align}\label{c4}
i\frac{\partial}{\partial s'}\hat{U}'_{j}+\frac{\partial^2}{\partial\tau'^2}\hat{U}'_{j}+2\left(\sum_{l=1,2}\hat{U}_{l}^{\prime\dag}\hat{U}'_{l}\right)\hat{U}'_{j}=R_j(\hat{U}'_{1},\hat{U}'_{2}).
\end{align}
where
\begin{eqnarray}\label{Rj}
&&R_j(\hat{U}'_{1},\hat{U}'_{2})=\notag\\
&&-2i\alpha_{j}s_{0}\hat{U}_{j}-i\frac{g_{\delta}s_{0}}{\tau_{0}}\frac{\partial}{\partial\tau'} \hat{U}'_{j}-2\beta_{j}\hat{U}_{3-j}^{\prime\dagger}\hat{U}'_{3-j}\hat{U}'_{j},
\end{eqnarray}
with $\beta_{j}=g_{j3-j}/g_{11}-1$. Note that,
under the DEIT condition, the absorption length $L_{j, A}$ and group velocity mismatch length $L_{\delta}$ are much larger than the dispersion length $L_{D}$, which means $\alpha_j=L_{D}/L_{j, A}\ll1$, and $g_{\delta}={\rm sgn}(\delta)L_{D}/L_{\delta}\ll1$. By choosing suitable system parameters, one can make $\beta_{j}\approx 0$ (i.e. $g_{j3-j}/g_{11}$). Thereby, the quantities $R_j(\hat{U}_{1},\hat{U}_{2})$ on the right-hand side of Eq.~(\ref{c1}) can be taken as perturbations. This can be realized if the magnetic field $B$ is not large and the two-photon detuning $\Delta_3$ is not far from the value $2.9\times10^{6}~{\rm Hz}$; see Fig.~\ref{Fig2} and the relevant statements in the main text. The situation where the influence of the perturbations $R_j$ (especially the external magnetic field $B$) plays a significant role is the work beyond the present study and will be considered elsewhere.

\section{Atomic spin squeezing for the class 3 defined by (\ref{class3})] }\label{app4}

In the class 1 and class 2 discussed in Sec.~\ref{APS}, we have assumed that there is no initial
coherence between the two atomic ground states $|1\rangle$ and $|2\rangle$, i.e. $S_{21}^{(0)}=S_{12}^{(0)}= 0$. If an initial coherence is prepared between the two ground states, the atomic ground state will be a coherent superposed state of $|1\rangle$ and $|2\rangle$, and hence one has
$S_{21}^{(0)}=S_{12}^{(0)}\neq 0$. In this situation, the system can also support significant atomic spin squeezing.

By using (\ref{class3}),
it is easy to show
\begin{align}
 \langle\hat{J}_{z}\rangle\neq0,\,\,\langle\hat{J}_{x}\rangle
 =\langle\hat{J}_{y}\rangle=0.
\end{align}
Based on the MHL equations (\ref{HLM0}) and (\ref{HLM1}),
for non-zero $S_{21}^{(0)}$ and $S_{12}^{(0)}$ we can obtain the solution
\begin{subequations}
\begin{align}
\hat{S}_{11}&\approx\hat{S}_{11}^{(0)}+a_{111}^{(2)}|g_{p}|^{2}\hat{E}_{p1}^{\dagger}\hat{E}_{p1},\\
\hat{S}_{22}&\approx\hat{S}_{22}^{(0)}+a_{222}^{(2)}|g_{p}|^{2}\hat{E}_{p2}^{\dagger}\hat{E}_{p2},\\
\hat{S}_{21}&\approx\hat{S}_{21}^{(0)}+a_{21}^{(2)}|g_{p}|^{2}
\hat{E}_{p2}^{\dagger}\hat{E}_{p1}.
\end{align}
\end{subequations}
This means that the dynamics of the atomic spins is similar to that of the probe-field polarization, and hence in this case the behavior of the atomic spin squeezing is similar to that of the polarization of the probe field.

\begin{figure}
\centering
\includegraphics[width=0.95\columnwidth]{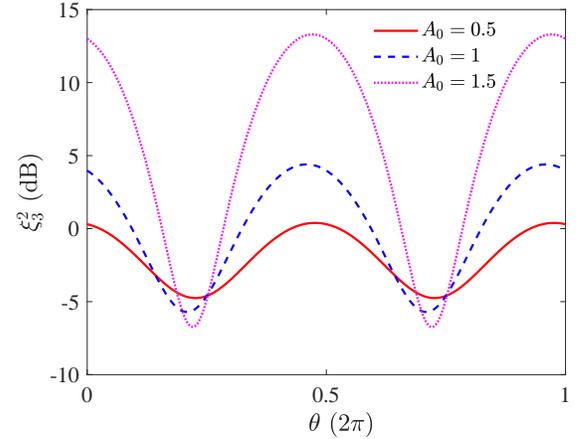}
\caption{
The degree of the spin squeezing $\xi^{2}_3$  (in dB) of the class 3 versus the detection angle $\theta$ at $s=0.5$,   for different dimensionless VOS amplitude $A_{0}=0.5$ (solid red line), 1 (dashed blue line) and 1.5 (dotted pink line), respectively. The initial coherence between the state $|1\rangle$ and $|2\rangle$ is assumed to be $S_{21}^{(0)}=S_{12}^{(0)}= 1/2$.
}
\label{Fig6}
\end{figure}

Shown in  Fig.~\ref{Fig6}
is the degree of the spin squeezing $\xi^{2}_3$  (in dB) of the spin class 3 as a function of the detection angle $\theta$ at $s=0.5$, respectively  for different dimensionless VOS amplitude $A_{0}=0.5$ (solid red line), 1 (dashed blue line) and 1.5 (dotted pink line). When plotting the figure, the initial coherence between the state $|1\rangle$ and $|2\rangle$ is chosen to be $S_{21}^{(0)}=S_{12}^{(0)}= 1/2$. We see that the behavior of $\xi_3^{2}$ is indeed similar to that of the polarization squeezing degree $R_{\theta}$ of the probe pulse [i.e. Fig.~\ref{Fig3}(b)]. The lower limit of $\xi^{2}_3$ for the degree of the atomic spin squeezing is $-6.1$\,dB.


\end{document}